\documentclass[fleqn,usenatbib]{mnras}

\usepackage{newtxtext,newtxmath}

\usepackage[T1]{fontenc}
\usepackage[dvipsnames]{xcolor}
\DeclareRobustCommand{\VAN}[3]{#2}
\let\VANthebibliography\thebibliography
\def\thebibliography{\DeclareRobustCommand{\VAN}[3]{##3}\VANthebibliography}

\usepackage{graphicx}	
\usepackage{amsmath}	
\usepackage{rotating}

\newcommand{\Msun}{\,M$_\odot$}
\newcommand{\Lsun}{\,L$_\odot$}
\newcommand{\Rsun}{\,R$_\odot$}
\newcommand{\tento}[1]{$10^{#1}$}
\newcommand{\timestento}[2]{$#1 \times 10^{#2}$}
\newcommand{\Msunyr}{\,$\mathrm{M}_{\odot}$\,$\mathrm{yr}^{-1}$}
 
\newcommand{\Mini}{${\rm M_{ini}}$}

\newcommand{\pols}{\citet{Pols:1998}\,}
\newcommand{\porb}{$P_{\rm orb}$}

\title[METISSE in BSE]{Modelling stellar evolution in mass-transferring binaries and gravitational-wave progenitors with METISSE}

\author[P. Agrawal et al.]
{Poojan Agrawal$^{1,3}$
, 
Jarrod Hurley$^{2,3}$,
Simon Stevenson$^{2,3}$, 
Carl L.~Rodriguez$^{1}$,
Dorottya Sz\'ecsi$^{4}$, 
\newauthor
Alex Kemp$^{5}$
\\
$^{1}$\,Department of Physics and Astronomy, The University of North Carolina at Chapel Hill, Chapel Hill, NC 27599, USA.\\
$^{2}$\,Centre for Astrophysics and Supercomputing, Swinburne University of Technology, Hawthorn, VIC 3122, Australia.\\
$^{3}$\,OzGrav: The ARC Centre of Excellence for Gravitational Wave Discovery, Hawthorn, VIC 3122, Australia.\\
$^{4}$\,Institute of Astronomy, Faculty of Physics, Astronomy and Informatics, Nicolaus Copernicus University, Grudziądzka 5, 87-100 Toruń, Poland.\\
$^{5}$\,Institute of Astronomy, KU Leuven, Celestijnenlaan 200 D, 3001 Leuven, Belgium.
}

\date{Accepted XXX. Received YYY; in original form ZZZ}

\pubyear{2023}

\begin{document}
\label{firstpage}
\pagerange{\pageref{firstpage}--\pageref{lastpage}}
\maketitle

\begin{abstract}

Massive binaries are vital sources of various transient processes, including gravitational-wave mergers.
However, large uncertainties in the evolution of massive stars, both physical and numerical, present a major challenge to the understanding of their binary evolution.
In this paper, we upgrade our interpolation-based stellar evolution code METISSE to include the effects of mass changes, such as binary mass transfer or wind-driven mass loss, not already included within the input stellar tracks.
METISSE's implementation of mass loss (applied to tracks without mass loss) shows excellent agreement with the SSE fitting formulae and with detailed MESA tracks, except in cases where the mass transfer is too rapid for the star to maintain equilibrium.
We use this updated version of METISSE within the binary population synthesis code BSE to demonstrate the impact of varying stellar evolution parameters, particularly core overshooting, on the evolution of a massive (25\Msun{} and 15\Msun{}) binary system with an orbital period of 1800\,days. 
Depending on the input tracks, we find that the binary system can form a binary black hole or a black hole-neutron star system, with primary\,(secondary) remnant masses ranging between 4.47\,(1.36)\Msun{} and 12.30\,(10.89)\Msun{}, and orbital periods ranging from 6\,days to the binary becoming unbound.
Extending this analysis to a population of isolated binaries uniformly distributed in mass and orbital period, we show that the input stellar models play an important role in determining which regions of the binary parameter space can produce compact binary mergers, paving the way for predictions for current and future gravitational-wave observatories.

\end{abstract}




\section{Introduction}

Stellar binaries play an important role in the evolution of the universe, opening up evolutionary pathways that would otherwise remain inaccessible through the evolution of single stars \citep[see e.g.,][]{Eldridge2020}. 
For massive stars, the role of binary evolution is even more crucial \citep{Sana:2012,deMink:2013} as massive binaries lead to the formation of X-ray binaries \citep{Verbunt1993}, gamma-ray bursts \citep{Woosley:2006}, kilo-novae \citep{Abbott2017c} and gravitational waves \citep[][]{Abbott:2016}, amongst many other astrophysical marvels. Recent studies even indicate the role of stellar triples in determining high-mass stellar evolution \citep[e.g.,][]{Eggleton:2008,Raghavan:2010,Moe:2017}.

With the release of the third LIGO-VIRGO-KAGRA gravitational-wave (GW) catalogue \citep[][]{GWTC3:2021}, we now have nearly 100 GW signals from the merger of binary systems of neutron stars (NSs) and black holes (BHs), the end states of massive stars. The parameters measured from GW signals, such as the spins and the masses of the binary components, help us determine the evolution of massive stars and their interaction with their neighbours \citep[e.g.,][]{Stevenson:2015,Mapelli:2017,Zevin:2017,Farmer:2020,Mandel:2022a}. The number of GW detections is expected to increase significantly in the future \citep{Abbott2018a}, and gravitational-wave astronomy is leading the way in shedding light onto the lives of massive stars. 
It is, therefore, a great time to incorporate the latest models of massive stars in our population synthesis codes and determine the impact of binarity on the evolution of stellar populations. 

Population synthesis codes serve as the tool to calculate key properties and interactions between stellar populations in galaxies and star clusters. 
However, keeping such codes up-to-date with the latest results of stellar evolution studies is an arduous task as many population synthesis codes rely on fitting formulae to approximate single star evolution.
These formulae are fast and robust but are not adaptable to the changes in the stellar tracks.

Advancements in computational ability has seen a large jump in the use of one-dimensional (1D) stellar evolution codes in the development of grids of massive binary star systems \citep[e.g.,][]{Langer:2020,Menon:2021,Sen:2022}. 
However, uncertainty in massive stellar evolution is not limited to the physical inputs but also depends on the numerical techniques employed by different stellar evolution codes for evolving such stars. 
As shown in \citet{Agrawal:2022b}, these numerical methods can have a non-trivial impact on the evolution of massive stars. 
It is, therefore, crucial to compare not just the tracks computed with different physical inputs but also the tracks that are evolved with different codes.

Recently, the method of interpolating between stellar tracks has gained popularity as it provides the flexibility of switching between sets of stellar tracks. 
Codes like SEVN \citep{Iorio:2022,Spera:2015} and COMBINE \citep{Kruckow:2018} make use of this method to compute single star properties for binary evolution. 
This helps study the effect of varying input physics on stellar tracks and hence on population synthesis models. 
Following a similar approach of interpolation, we have developed MEthod of Interpolation for Single Star Evolution \citep[METISSE;][]{Agrawal:2020} as an alternative to the Single Star Evolution \citep[SSE;][]{Hurley:2000} fitting formulae in population synthesis codes. 
METISSE interpolates between sets of pre-computed stellar tracks to approximate evolution parameters for a population of stars. 
Similar to other interpolation-based rapid codes, METISSE can readily make use of stellar models computed with different stellar evolution codes. 

In this work, we update METISSE to include the implications of additional mass changes, such as mass loss/gain through stellar winds, or mass loss/gain through mass transfer, that are not present in the underlying detailed stellar evolution tracks used as input. 
We combine this updated version of METISSE with the Binary Stellar Evolution \citep[BSE;][]{Hurley:2002} code. BSE is a popular binary population synthesis code that can rapidly compute the evolution of stars in a binary system, along with providing algorithms to model the necessary binary physics. 
BSE relies on the subroutines and the fitting formulae from SSE to compute single stellar evolution properties. The resulting algorithm is computationally cheap, fast and robust and many other binary evolution codes such as STARTRACK \citep{Belczynski:2002,Belczynski:2008}, binary\_c \citep{Izzard2004,Claeys2014}, COMPAS \citep{Stevenson:2017,VignaGomez:2018}, MOBSE \citep{Giacobbo2018} and COSMIC \citep{Breivik2020} are based on BSE.

This study lays the groundwork needed to combine our most up-to-date understanding of massive stars with population synthesis codes to constrain the formation of GW progenitors in stellar multiples and star clusters. 
We briefly summarise the capabilities and method of METISSE as a single stellar evolution code in Section~\ref{sec:metisse_as_sse}. 
We describe our implementation of extra mass loss in METISSE in Section~\ref{sec:metisse_mt}.
In Section~\ref{sec:results_metisse_mt} we test the validity of the implementation by comparing the stellar tracks interpolated by METISSE with additional mass loss to the results from SSE and from the 1D stellar evolution code Modules for Experiments in Stellar Astrophysics \citep[MESA;][]{Paxton2019}. 
In Section~\ref{sec:case_study}, we present a case study of a 25\Msun{} and 15\Msun{} binary evolved using SSE in BSE (BSE-SSE) and compare the results with METISSE in BSE (BSE-METISSE) using stellar evolution tracks from \pols and three newly computed MESA tracks that differ only in core overshooting.
We extend this analysis to a population of isolated binaries in Section~\ref{sec:binary_pop}.
We present our conclusions and discuss potential future work in Section~\ref{sec:conclusions_chap4}.

\begin{table*}
\begin{tabular*}{\textwidth}{|l|l|l|}
\hline
NAME & FUNCTION IN SSE & FUNCTION IN METISSE\\
\hline
\textsc{ZCNSTS} & Set all the constants of the formulae which depend on metallicity & Read detailed models of given metallicity and find mass cutoffs\\

\textsc{STAR} & Derive the parameters that divide the various evolution stages & Interpolate between detailed models to get a track of given mass\\

\textsc{HRDIAG} & Decide which evolution stage the star is currently at, and calculate & Interpolate within the new track to determine stellar parameters at given
\\
& the appropriate luminosity, radius and core mass &  age\\
\textsc{MLWIND} & Derive the mass loss as a function of evolution stage & Same as SSE\\

\textsc{DELTAT} & Calculate the time-steps depending on the stage of evolution & Same as SSE\\
\hline
\end{tabular*}
\caption{Major subroutines in the SSE and METISSE packages and their functionality in each.}
\label{tab:sse_subroutines}
\end{table*}

\section{METISSE as a Single Star Evolution Code}
\label{sec:metisse_as_sse}

METISSE is a synthetic stellar evolution code that can quickly compute the evolution of many stars by interpolating between a finite set of models computed with 1D stellar evolution codes (also known as detailed stellar evolution codes). 
METISSE has been designed to serve as an alternative to the SSE fitting formulae in population modelling codes such as BSE and NBODY6 \citep{Aarseth2003}. 
However it can also be used as a stand-alone code, for example, for population synthesis of single stars, or to test input stellar tracks. 
Further capabilities of METISSE are described in \citet{Agrawal:2020}.

The SSE package consists of several subroutines (sub-program units in Fortran), where each subroutine has a particular role. 
These subroutines are called by an overarching subroutine called \textsc{Evolv1} for calculating the evolution of a single star. 
For computing the evolution of a binary star in BSE, the subroutines in Table \ref{tab:sse_subroutines} are called twice by a similar overlying subprogram \textsc{Evolv2} to compute parameters of two stars at every time-step of the evolution and calculate relevant binary evolution parameters. 
An exception to this is the subroutine \textsc{Zcnsts} which needs to be called separately at the beginning of the program for calculations dependent on the metallicity.

METISSE is structured to contain similar subroutines to SSE with the same functionality. 
Major SSE subroutines, their functions and their equivalent in METISSE are listed in Table \ref{tab:sse_subroutines}. 
These subroutines mimic the behaviour of the SSE subroutines from the outside, with the same name and input/output variables. 
They have been written in FORTRAN90/95 and make use of the modern Fortran architecture for efficiently storing and passing large arrays of data required for interpolation. 

Before explaining the implementation of mass transfer, we briefly summarize the usual interpolation process in METISSE. 
For use in METISSE, input tracks must be divided into equivalent evolutionary phases \citep[][]{Prather:1976PhDT,Bergbusch:2001}.
These phases are readily identifiable by evolutionary features such as the central hydrogen mass fraction, with examples being the main-sequence (MS) phase or the Hertzsprung Gap (HG) phase.
Each evolutionary phase is further subdivided into an equally spaced set of points called Equivalent Evolutionary Points (EEPs), fixed in number across all masses. 
For a given initial mass, \Mini, an evolutionary track is calculated by interpolating between the corresponding EEPs of the neighbouring mass tracks. The type of interpolation performed is either linear or monotonic piece-wise cubic interpolation \citep{Steffen:1990}, depending on the number of tracks available in the neighbourhood of \Mini.
The resulting track is a collection of stellar parameters at each EEP.
METISSE further interpolates within the mass-interpolated track, between the EEPs enveloping the given age, to calculate stellar parameters at any instant. 

\section{Implementing mass transfer in METISSE}
\label{sec:metisse_mt} 

The total mass of a star can change due to mass loss through stellar winds or due to mass transfer during interaction with a binary companion. 
Depending on the evolutionary phase, both mass loss and gain can have a substantial impact on the structure and the evolution of the star. 
Hence, it is important to incorporate the effects of mass change in the stellar tracks.

The stellar models from \citet{Pols:1998} that were used in calculating the SSE fitting formulae did not include mass loss. 
A change of stellar parameters in response to any kind of mass change is dealt with by making use of the current mass of the star $M_t$ and the \textit{`effective'} initial mass of the star $M_0$ (which can be distinct from the zero-age main-sequence -- ZAMS -- mass of the star). 
Parameters such as luminosity, timescales and core mass are calculated using $M_0$ while the radius is calculated using $M_t$ in their respective formulae. 
On the main sequence, it is assumed that $M_0$ is equal to $M_t$ and in response to any mass change, $M_0$ is adjusted to account for the corresponding change to the main-sequence lifetime. 

In METISSE, if wind mass loss is already incorporated in the input stellar models computed with a detailed code, one can simply interpolate between them to achieve the same effect. 
It is, in fact, more accurate compared to fitting formulae as the changes to stellar structure due to mass loss are better modelled in 1D stellar codes, a benefit which is directly carried over into METISSE. 
However, we still need to take into account the impact of mass changes resulting from interaction with a binary companion, as well as the case where input tracks computed without any mass loss are used. 

In the absence of any extra changes in mass, mass interpolation only happens once for a star of the given initial mass. If for any reason the mass does change and the change, $dm = M_{\rm t,prev}-M_t$, exceeds \tento{-6}\Msun{} at any time-step (due to either mass loss or mass gain), a new mass-interpolated track is calculated before proceeding with the age interpolation to determine the stellar parameters for the age at that step. 
The procedure for calculating the new track is described below, where the symbol prime $(')$ indicates the properties of the star at time $t$ after the mass change has been taken into account.

For a mass change $dm$ at a time $t$ in a star's life,
\begin{itemize}
    \item The first step is to locate the nearest EEP to $t$, say ${\rm EEP}_i$, in the current track. Since the evolutionary parameters of the input stellar tracks and the track corresponding to the current mass of the star are only stored at the EEPs, the mass change is applied to the total mass of the star at ${\rm EEP}_i$  ($M_{\rm EEP_i}$) to get $M'_{\rm EEP_i} = M_{\rm EEP_i}-dm$.
    
    \item Next, METISSE searches all input tracks to find the two tracks whose masses at ${\rm EEP}_i$ envelop $M'_{\rm EEP_i}$. It then linearly interpolates between their initial masses to get the initial mass ($M'_{\rm ini}$) of the star whose mass at ${\rm EEP}_i$ will be $M'_{\rm EEP_i}$. 

    \item Finally we interpolate the new track, using the new initial mass ($M'_{\rm ini}$) and the method described in Section~\ref{sec:metisse_as_sse}. 
    
\end{itemize}

For input tracks computed without any mass loss, the new initial mass, $M'_{\rm ini}$ and $M'_{\rm EEP_i}$ in METISSE are the same as the effective initial mass, $M_0$ and $M_t$ of SSE.  
To account for the effect of mass loss on different stellar parameters, we currently apply similar assumptions in METISSE as in SSE, i.e., stellar parameters are designated to be dependent on either the track corresponding to the current mass $M_t$ or the track corresponding to the effective initial mass of the star $M_0$. 

The calculation of parameters beyond $t$ depends on the evolutionary phase of the star.
On the main sequence, the stellar parameters are highly sensitive to the total mass of the star, therefore all parameters of the star beyond time $t$ are calculated using the new track. 
However, a lower-mass star (assuming mass loss) at the same age would have burned less hydrogen. 
Therefore, we follow SSE, and age the star to conserve the fraction of hydrogen burned. 
The new age $t'$ is calculated from the actual age of the star $t$, using the main-sequence time for the old track ($t_{\rm MS}$) and the newly interpolated track ($t'_{\rm MS}$) \citep[cf.][]{Hurley:2000} as follows,

\begin{equation}
     \centering
    t' = \frac{t'_{\rm MS}}{t_{\rm MS}}t \, .
\end{equation}

During the post-main-sequence evolution of the star, the core evolution is assumed to be unaffected by changes to the total mass. Therefore, parameters related to the core such as the core mass and luminosity are calculated using the actual age $t$ and the track corresponding to the mass of the star at the end of the main sequence, i.e., the track where $M_{\rm ini}=M_0$. 
Other parameters, such as the stellar radius, are calculated using the age $t'$ and the new track corresponding to the current total mass of the star. 

The new age $t'$ is calculated such that the fractional time spent in the current phase is the same in both tracks.
Therefore, if $t_i$ is the time at the beginning of the phase and $(t_{\rm phase})$ is the duration of the current phase for the track corresponding to $M_{\rm ini}=M_0$, then,

\begin{equation}
     \centering
    t' = t'_{\rm start} + \frac{(t-t_{\rm start})}{t_{\rm phase}}t'_{\rm phase} \, .
    \label{eq:t_post_ms}
\end{equation}

The above procedure allows surface parameters to change according to the new total mass of the star while conserving the core properties.

For massive stars ($M_{\rm ini} > 10$\Msun{}), mass loss due to stellar winds can be quite high ($\sim$\tento{-4}\Msun{}yr$^{-1}$) and the star can lose its envelope before nuclear burning is completed \citep[e.g.,][]{Pols2002,McClelland:2016,Woosley2019,Laplace2020}. Depending on its mass and the evolutionary stage, a stripped star in METISSE can either become a helium white dwarf or a naked helium star. We currently revert to using SSE formulae for further evolution of the naked helium stars. 

\begin{figure}
	\includegraphics[width=\columnwidth]{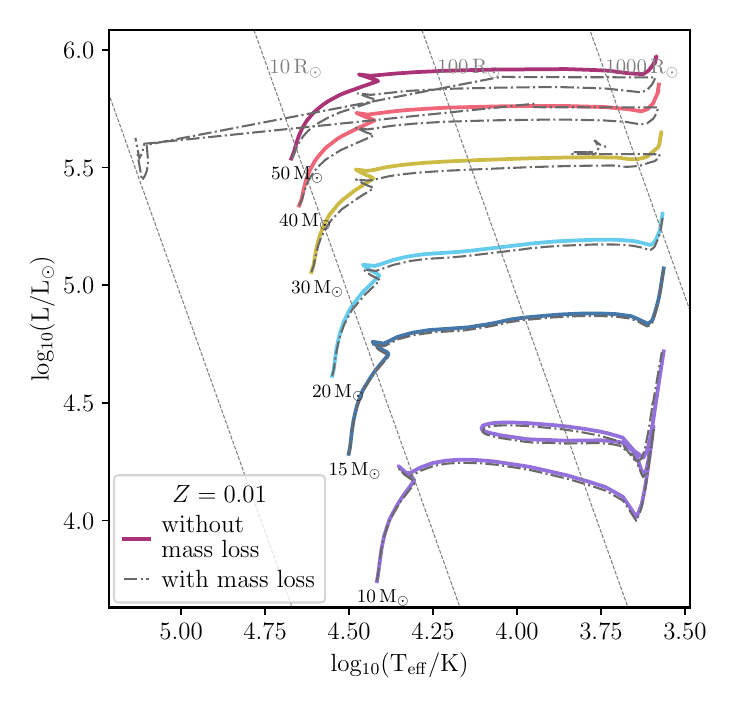}
    \caption{Hertzsprung–Russell (HR) diagram showing tracks interpolated by METISSE using detailed tracks from \citet{Pols:1998} for metallicity, $Z=$ \tento{-2}. Solid lines represent tracks without any mass loss while dashed lines represent the tracks including wind mass loss as described in Section~\ref{sec:results_metisse_mt}.}
    \label{fig:metisse_ml}
\end{figure}

\begin{figure}
    \includegraphics[width=\columnwidth]{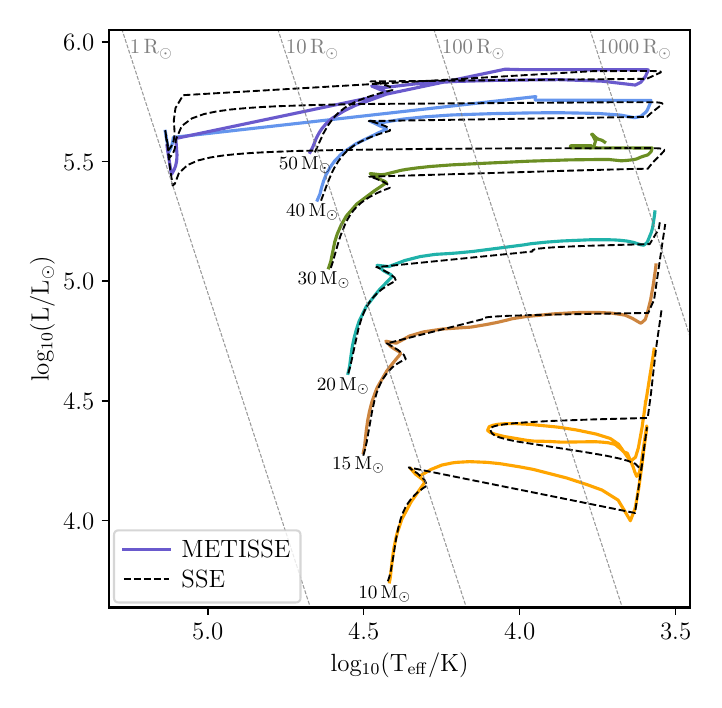}
    \caption{HR diagram comparing stellar tracks with mass loss as computed by SSE and METISSE using \citet{Pols:1998} models for $Z=$ \tento{-2}. For both sets wind mass-loss rates from \citet{Belczynski:2010} were used in computing the tracks.  
    }
    \label{fig:hrd_sse_metisse_ml}
\end{figure}

\section{Testing the validity of the implementation}
\label{sec:results_metisse_mt}

In this section, we test the validity of the mass-loss implementation in METISSE described in Section~\ref{sec:metisse_mt}. 
To calculate mass-loss rates across different evolutionary phases, we follow the algorithm described by \citet{Belczynski:2010}, which is the latest mass-loss algorithm available in SSE.
We apply these mass-loss rates consistently across all models we consider in this paper, including in METISSE to stellar tracks without any mass loss (both those from \citealp{Pols:1998} and those from MESA), in SSE (which is also based on \citet{Pols:1998} models without mass loss) and in MESA for computing detailed models with mass loss.
Below we briefly describe the different components of the mass-loss scheme. 

If the surface luminosity $(L)$ of the star exceeds $10^{5}$\Lsun{} and the radius $(R)$ satisfies, $(R/$\Rsun $)\,(L/$\Lsun$)^{0.5}>10^{5}$, a fixed mass loss of \timestento{1.5}{-4}\Msunyr{} is applied \citep[see equation 8 of][]{Belczynski:2010}, to account for Luminous Blue Variable (LBV) behaviour of the stars. For stars whose effective surface temperature, $T_{\rm {eff}}$, is in the range of 12500\,K and 50000\,K, mass-loss rates from \citet{Vink:2000, Vink:2001} are used with the iron bi-stability jump temperature at 25000K. 
For stars not covered by the above $T_{\rm {eff}}$ criteria but that are more luminous than 4000\Lsun{}, the mass-loss rates from \citet{NieuwenhuijzenanddeJager:1990} are applied with a metallicity correction factor of $Z^{0.5}$ from \citet{Kudritzki:1989}.

Where none of the above conditions is met, i.e., for low-mass stars, wind mass-loss rates from \citet{Reimers:1978} are used on the giant branch (GB) and from \citet{VassiliadisWood:1993} on the asymptotic giant branch (AGB), including the thermally-pulsating AGB (TPAGB).
For naked helium stars, and for stars with small hydrogen-rich fractional envelope mass \citep[$\mu$; see equation 97 of][]{Hurley:2000} the wind prescription from \citet{Hamann:1998} is used (reduced by a factor $1-\mu$ in the latter case) with a metallicity scaling factor of $Z^{0.86}$ from \citet{Vink:2005}. 
We use $Z_\odot=$ \timestento{2}{-2} as the reference solar metallicity in calculating the mass-loss rates as they have originally been scaled from this value. 

Fig\,\ref{fig:metisse_ml} shows the evolutionary tracks, computed with and without mass loss with METISSE using \citet{Pols:1998} models at metallicity $Z=$ \tento{-2}. 
The effect of mass loss is visible on the stars more massive than 10\Msun{} in the Hertzsprung–Russell (HR) diagram. 
Tracks with mass loss are less luminous than tracks without mass loss. 
For the 40\Msun{} track, the inclusion of mass loss leads to a completely different final state, i.e., a naked helium star when the mass loss is included, compared to a red supergiant when not including any mass loss. 
At the end of its life, the 40\Msun{} star without any mass loss forms a 36\Msun{} BH while the same star with mass loss forms just an 11.8\Msun{} BH. 
This simple comparison highlights the impact of mass loss on stellar evolution, especially for massive stars. 


\subsection{Comparing tracks with mass loss for METISSE and SSE}
\label{subsec:metisse_sse}

In this section, we compare the stellar tracks given by the fitting formulae of SSE with the tracks interpolated by METISSE with mass loss using the \citet{Pols:1998} models in the mass range 0.5--50\Msun{} at  $Z=$ \tento{-2} metallicity.
We exclude stars less massive than 10\Msun{} in our comparison here, as wind mass-loss rates are too small for them to cause any significant difference. 

\subsubsection{Differences in the HR diagram}
\label{sec:hrd_metisse_sse}

Figure~\ref{fig:hrd_sse_metisse_ml} shows the comparison between the evolutionary tracks computed by METISSE and SSE.
Inspection of the figure reveals that the tracks agree reasonably well during the main-sequence phase but start diverging during the post-main-sequence evolution. 
As shown in \citet{Agrawal:2020}, METISSE is better at reproducing input stellar tracks than the fitting formulae from SSE. 
The discrepancy between the methods is the main source of variation in the 10\Msun{} track in Figure~\ref{fig:hrd_sse_metisse_ml}. 
For stars with initial masses 20\Msun{} and above, the maximum difference occurs towards the end of core helium burning where METISSE systematically predicts higher $T_{\rm eff}$ than SSE.

\begin{figure}
    \includegraphics[width=\columnwidth]{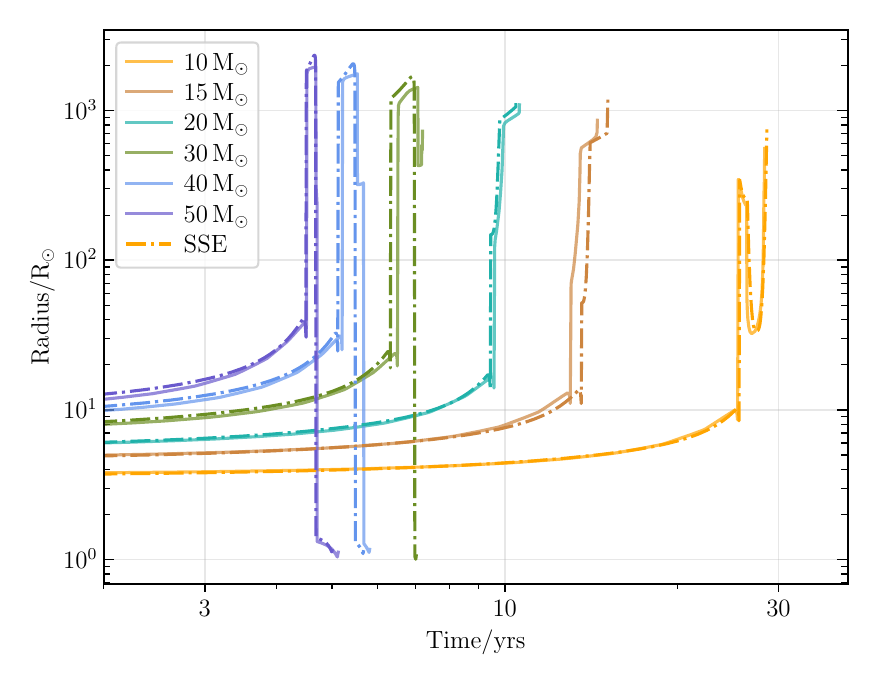}
    \caption{Radius versus time for stellar tracks computed with mass loss using METISSE (solid lines) with \citet{Pols:1998} models and using SSE (dashed-dotted lines) for $Z=$ \tento{-2}. For stars more massive than 15\Msun, the radial evolution agrees for most of the evolution except beyond core helium burning.}
    \label{fig:rad_sse_metisse}
\end{figure}

To further understand the origin of these differences, we compare the radial evolution of the stars as predicted by METISSE and SSE in Figure~\ref{fig:rad_sse_metisse}. Similar to Figure~\ref{fig:hrd_sse_metisse_ml}, results from METISSE show an overall agreement with SSE. 
An exception is the 15\Msun{} star that shows slight disagreement in the time of evolution.
This is due to the proximity of the effective initial mass of the 15\Msun{} star to one of the critical masses at 16\Msun{}. The critical masses in METISSE serve as the lower limits above which certain physical properties start to appear for stellar tracks and interpolation between these tracks can lead to physically incorrect results \citep[cf.][]{Agrawal:2020}. 
Therefore, at each step of mass loss, the intermediate tracks during the evolution of the 15\Msun{} star in METISSE have been extrapolated from the 16\Msun{} and 20\Msun{} models, leading to a slight deviation from a 15\Msun{} track computed with SSE.

For other high-mass stars (defined as the stars that ignite helium on the HG and do not undergo the RGB phase, stars more massive than 10\Msun{} here), SSE predicts a monotonic increase in radii during the core-helium-burning phase unless they lose their envelope and become naked helium stars. For the same phase of evolution, METISSE predicts a comparatively lower value of the maximum radial expansion that these stars can achieve. 
A substantial effect of the lower radii predicted by METISSE for the post-main-sequence evolution is on the evolution of a 30\Msun{} star.
While SSE predicts that the 30\Msun{} star will become a naked helium star, METISSE predicts the formation of a supergiant at the end of core helium burning. 

This discrepancy towards the end of the core helium burning phase for high-mass stars arises from technical details relating to how mass loss is implemented for this phase in both codes. 
In SSE, the radius calculation for high-mass stars transitions to using the fitting formulae for the AGB phase after a certain fraction of time (based on the mass fraction of the convective envelope from a fit to the \citealt{Pols:1998} tracks) during the core helium burning phase \citep[see Section 5.3 of][]{Hurley:2000}. 
By doing so, SSE forces the star to start ascending the AGB even before the end of core helium burning. 
However, high mass-loss rates can strip the star of its envelope before it can reach the AGB after core helium burning. 
In such cases stellar radii predicted by SSE are larger than their actual value. 

As shown by recent 1D stellar models, these high-mass stars develop sub-surface convective layers during core helium burning as they evolve towards the giant branch. 
These layers can readjust thermally and stellar models only show a modest change in their radius in response to mass loss \citep[e.g., see][]{Woods:2011,Passy:2012}. 
Therefore, in METISSE, we do not impose the condition that the star will ascend the giant branch at the end of core helium burning. 
Instead, the values of the stellar radius from the previous mass-interpolated track are retained if the mass of the convective envelope exceeds 20 per cent of the total mass of the star at any time-step in METISSE. 

\begin{figure}
\includegraphics[width=\columnwidth]{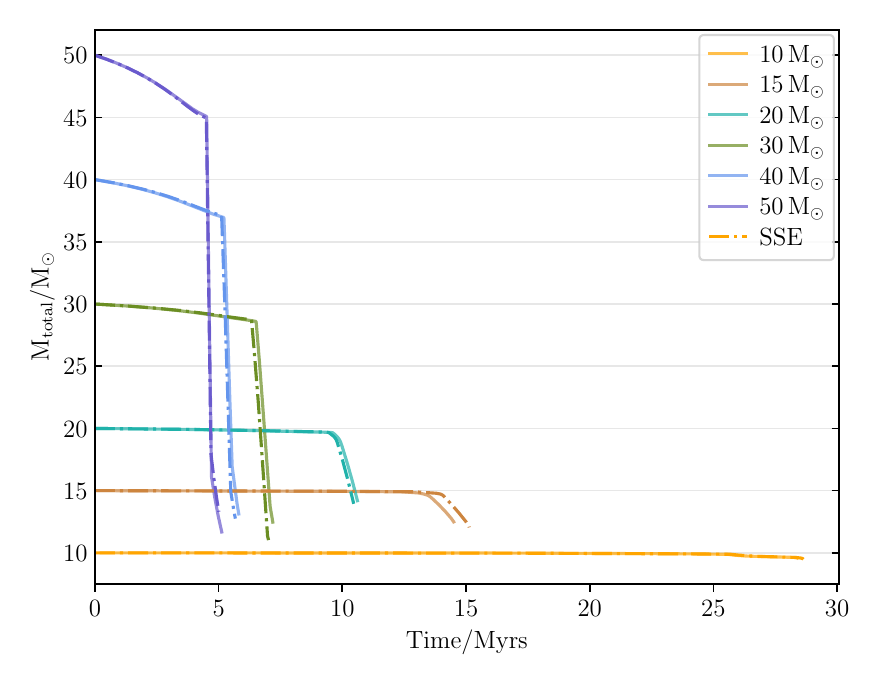}
\caption{Evolution of total mass with time for stars computed with mass loss by METISSE (solid lines) and SSE (dashed-dotted lines). During the late evolution of the 20--40\Msun{} stars, METISSE predicts a shallower decrease in mass compared to SSE. This agrees with the behaviour of stellar tracks in the HRD in Figure~\ref{fig:hrd_sse_metisse_ml}.}
\label{fig:Mass_sse_metisse_ml}
\end{figure}

 \begin{figure}
	\includegraphics[width=\columnwidth]{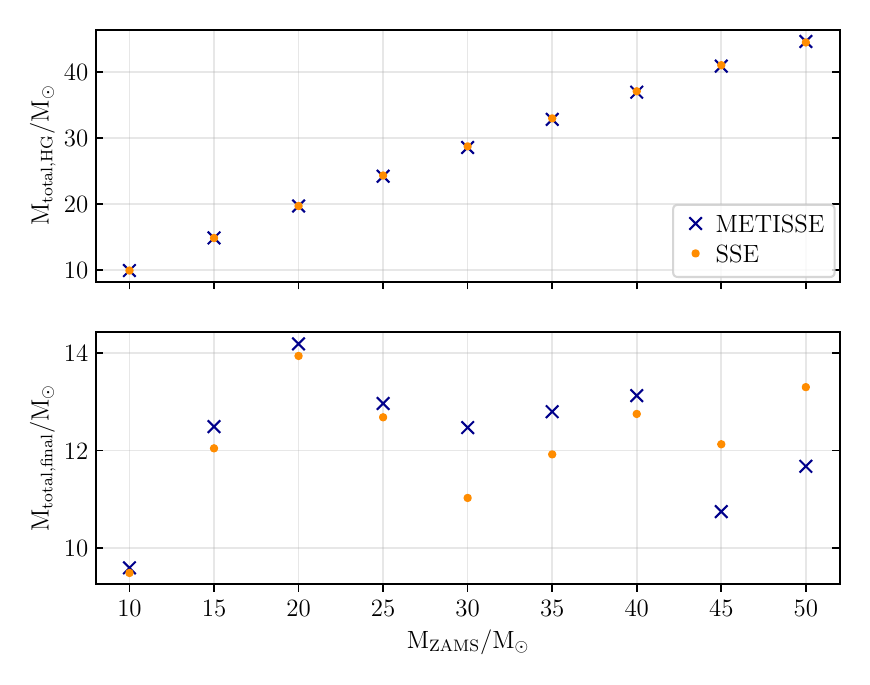}
    \caption{Total mass of the star as a function of its Zero-Age Main-Sequence (ZAMS) mass as predicted by SSE (orange dots) and METISSE (blue crosses). The top panel represents the total mass at the end of the HG (when core helium burning begins) while the bottom panel shows the values at the end of the nuclear-burning life of the star (before the star becomes a remnant).}
    \label{fig:Mend_sse_metisse_ml}
\end{figure}

\begin{figure}
    \includegraphics[width=\columnwidth]{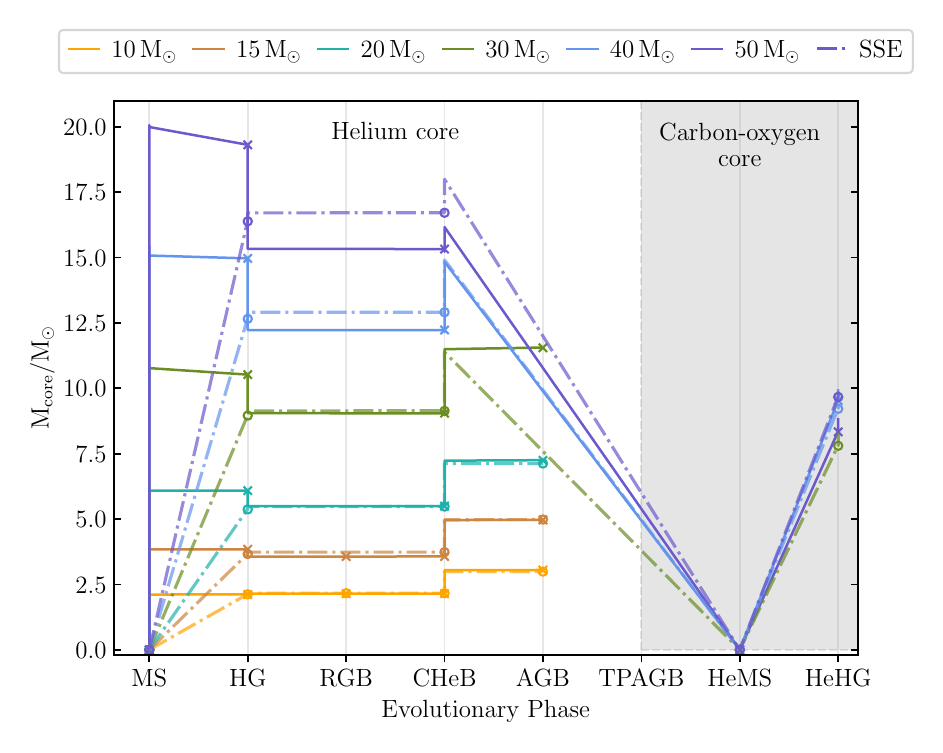}
    \caption{Evolution of core mass with phase as predicted by SSE (dashed-dotted) and METISSE (solid lines). Circles (METISSE) and crosses (SSE) denote the core mass at the beginning of each evolutionary phase.
    Following the SSE convention, $M_{\rm core}$ here represents the mass of the helium-rich core for phases up to the AGB and the mass of the carbon-oxygen core for phases beyond the AGB (including naked helium star phases). }
    \label{fig:core_mass_sse_metisse_ml}
\end{figure}

\begin{figure*}
\begin{tabular}{cc}
\includegraphics[width=\columnwidth]{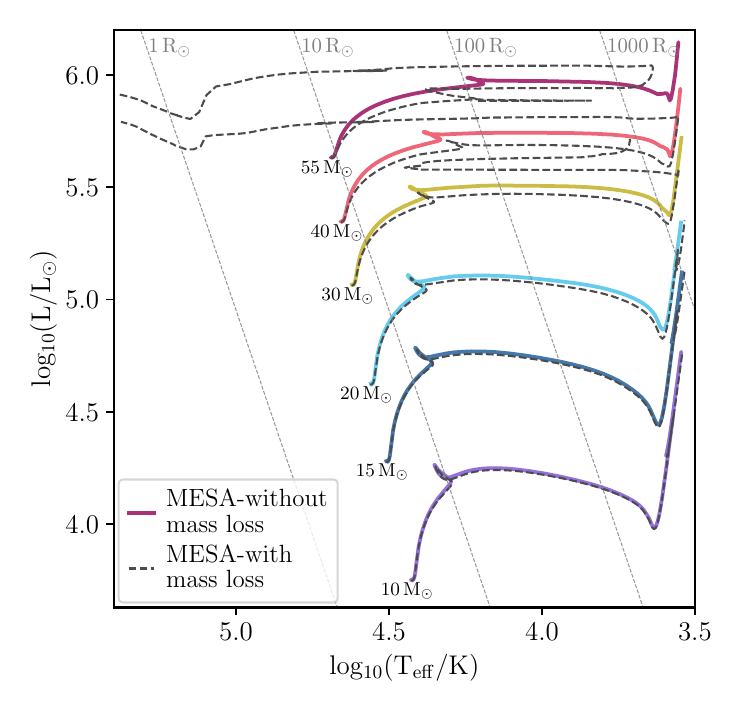}
&
\includegraphics[width=\columnwidth]{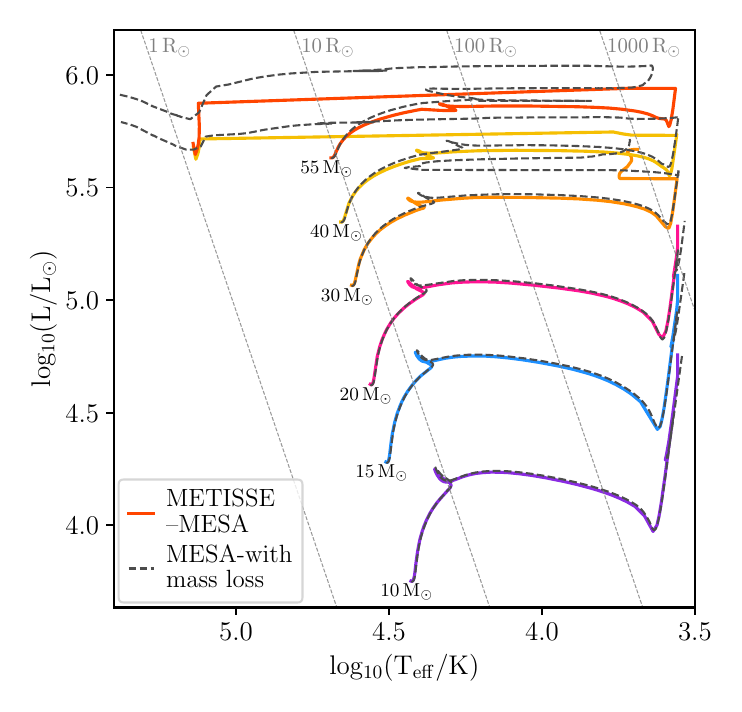}
\\
\end{tabular}

\caption{
The left panel shows the evolutionary tracks computed using MESA for metallicity $Z=$ \tento{-2} without any mass loss (solid lines) and with mass-loss rates from \citet{Belczynski:2010} (dashed lines). For the same metallicity, the right panel presents the comparison between stellar tracks interpolated by METISSE (solid lines) using stellar models without mass loss from MESA (adding mass loss on top) with the detailed models from MESA (dashed lines), computed using the same mass-loss rates as in METISSE. While METISSE shows good agreement with MESA at low mass-loss rates, it predicts lower luminosity for high mass-loss rates on the main sequence. }
\label{fig:hrd_metisse_mesa_Zsun}
\end{figure*}

\begin{figure}
\begin{tabular}{c}
     \includegraphics[width=\columnwidth]{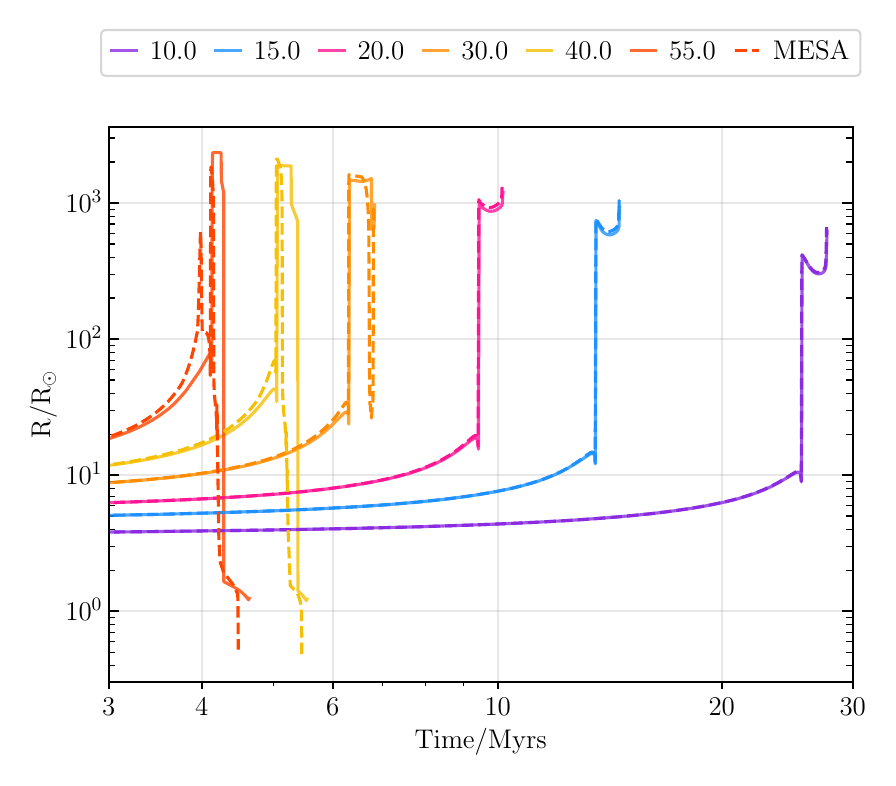}
    \\
    \includegraphics[width=\columnwidth]{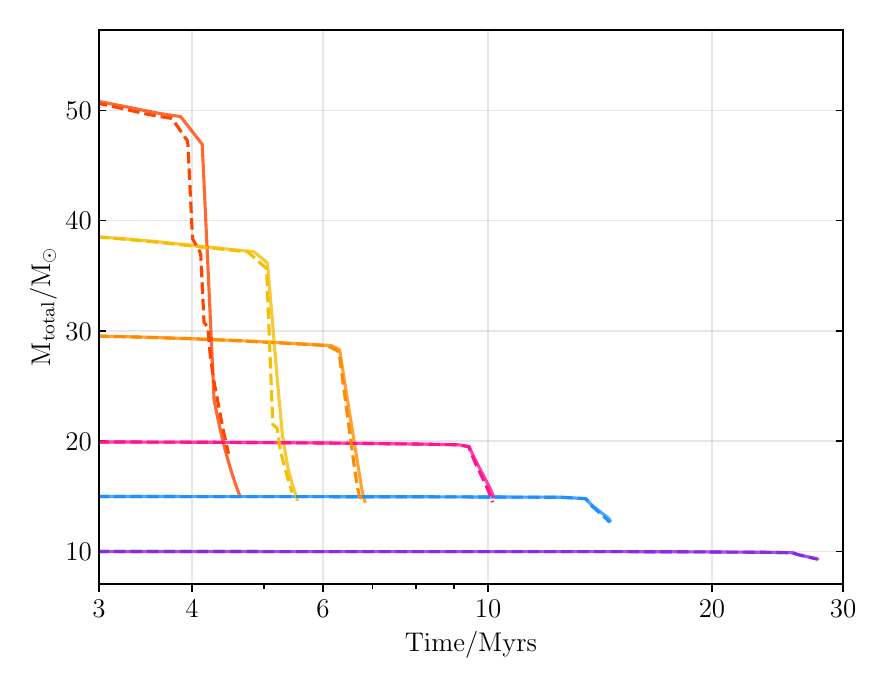}\\
    \includegraphics[width=\columnwidth]{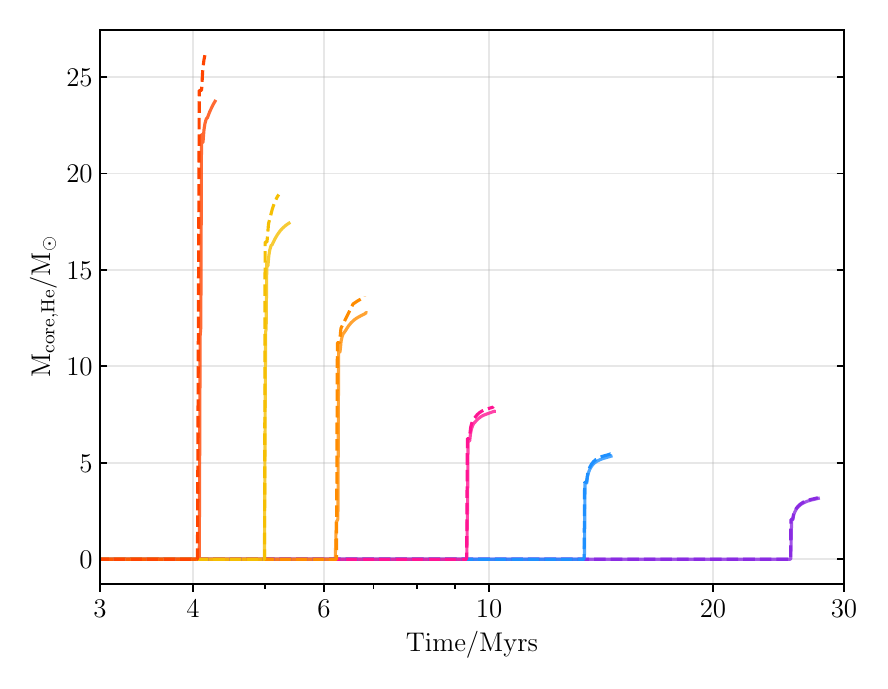}  \\
\end{tabular}

\caption{Evolution of stellar radius (top panel), total mass (middle panel) and helium core mass (bottom panel), as predicted by METISSE (in solid lines) and MESA (in dashed lines), for stars with $Z=$ \tento{-2}. Similar to the HRD in Figure~\ref{fig:hrd_metisse_mesa_Zsun}, METISSE can closely reproduce the results from MESA except at high mass-loss rates on the main sequence, where METISSE predicts lower values compared to MESA.}

\label{fig:mass_metisse_mesa}
\end{figure}

\subsubsection{Differences in the evolution of total mass with time}
\label{subsubsec:diff_evol_total_mass_time}

In Figure~\ref{fig:Mass_sse_metisse_ml}, we show the total mass of the star with time as computed by METISSE and SSE. The evolution of total mass for the 10\Msun{} track matches very well between the codes. For more massive stars, the decrease of mass with time is quite similar for most of their evolution. Differences appear only in the last $\sim$1\,Myr, where SSE predicts a steeper decrease in mass compared to METISSE.  
To better understand the origin of these differences, we plot the total mass of the star as a function of its zero-age main-sequence mass at the end of different evolutionary phases for both SSE and METISSE in Figure~\ref{fig:Mend_sse_metisse_ml}. We find that the total mass of the stars agrees well until core helium burning begins (at the end of the HG) but starts to differ thereafter. By the end of nuclear burning (before the star becomes a remnant), the maximum predicted difference can be up to 1.5\Msun{}, and happens for 25 and 30\Msun{} stars. 
The variations in mass are similar to the variations in the stellar tracks in the HR diagram during core helium burning.
Since mass-loss rates depend on the surface properties of the star, particularly radius, these differences can be associated with the differences that we see in the tracks in the HR diagram. 

\subsubsection{Differences in the evolution of core mass with phase}
\label{subsec:Mcore_sse_metisse}

As a star loses its envelope owing to stellar winds or binary mass transfer, the role of the core becomes increasingly important in determining stellar properties, especially the final fate of the star. 
There are several ways for determining the core boundaries in the detailed models based on properties such as chemical composition and sound speed (compressibility) \citep[for example, see][]{Ivanova:2011,Vigna-Gomez:2022}. 
Usually, the core is defined as the region interior to the boundary depleted in the element(s) undergoing nuclear fusion in the core and rich in the nuclear product(s). 
Core boundaries are not fixed but can increase e.g., as shell burning adds processed material, or even decrease e.g., as convection, dredge-ups and other mixing events dilute the outer layers of the core with the unprocessed material from the envelope. 

In Figure~\ref{fig:core_mass_sse_metisse_ml} we show the core mass of the star with evolutionary phase for stellar tracks computed by METISSE and SSE. 
Following SSE's definition, core mass in the figure represents the mass the dominant core, which is the helium-rich core for phases up to the AGB and the carbon-oxygen core for phases beyond the AGB.

While both SSE and METISSE predict that all stars start with zero core mass at the ZAMS, core mass in SSE is assumed to be zero throughout MS and only becomes non-zero at the beginning of the HG.
However, in 1-D stellar models, a core is usually defined according to mass-fraction thresholds. Thus a star can have well defined helium core before it reaches the HG. 
Since METISSE completely relies on the data from the underlying stellar models, it predicts non-zero helium core masses towards the end of MS phase.

For other evolutionary phases, the core masses at the end of each evolutionary phase for stars up to 20\Msun{} show a good agreement between METISSE and SSE, with the maximum difference being 0.2\Msun{} for a 15\Msun{} star at the end of the HG.
Moreover, METISSE is better able to reflect the variation in core mass with time (due to mixing, shell burning etc.) compared to SSE where the fitting formulae are designed to only register the increase in core mass.

For stars with initial masses less than 30\Msun{}, stellar winds are not strong enough to strip them of their hydrogen-rich envelope, and these stars end their life as supergiants on the AGB phase. However, more massive stars (stars with initial masses of 40\Msun{} and above) lose their envelopes owing to high wind mass-loss rates and become naked helium stars. 
According to SSE, when a star loses its envelope during CHeB, it becomes a HeMS star with zero (carbon-oxygen) core mass. In METISSE, we currently switch to using SSE formulae when a star loses its envelope. Therefore it also predicts zero core mass for stars during HeMS phase. However, differences in predictions of the total mass of the star by SSE and METISSE (Figure~\ref{fig:Mass_sse_metisse_ml} and Figure~\ref{fig:Mend_sse_metisse_ml}) lead to variation in the carbon-oxygen core mass predictions during HeHG.

For stars with initial masses of 40\Msun{} and above, where both SSE and METISSE predict the formation of a naked helium star, there can be a difference of up to 1.4\Msun{} in the core masses (helium core) during the HG and between the two codes. This eventuates in the difference of up to 1.25\Msun{} in the carbon-oxygen core by the end of star's life.
For the 30\Msun{} star, the difference in the core mass prediction at the end of HG is less than 0.1\Msun{}. 
However, lower radii predicted by METISSE during core helium burning causes a shallower decrease in mass. 
Therefore, despite having a similar envelope mass (19.50\Msun{}) compared to SSE (19.56\Msun{}) at the beginning of core helium burning, the 30\Msun{} star in METISSE is able to retain its envelope and end its life as a red supergiant. 
The difference in the evolutionary paths for the 30\Msun{} star beyond core helium burning shows up as a difference of about 3\Msun{} in the core mass at the end of nuclear burning. 
It is important to highlight that this difference is between the helium core mass in METISSE and the carbon-oxygen core mass in SSE. The difference between the respective carbon-oxygen core masses is only 0.2\Msun{}.

\subsection{Comparing METISSE with detailed models with mass loss from MESA}
\label{subsec:metisse_mesa}

Any method of rapid population synthesis should be able to mimic the results of the detailed evolution as closely as possible. 
The method for accounting for mass changes in stellar tracks in METISSE shows good agreement with the results of SSE. 
However, the ideal scenario will be to reproduce results from detailed evolution computed including mass loss. 
Therefore, in this section, we test the validity of our method using models computed from MESA both with and without mass loss. 

For this purpose, we have computed two sets of models with MESA for stars in the mass range of 8--55\Msun{} at metallicity $Z=$ \tento{-2}. Both sets of models employ the basic 21 isotope nuclear reaction network of MESA and have been computed through to the end of carbon burning in the core. Other physical inputs are described in \citet{Agrawal:2022b}.

The first set of models with MESA have been computed without any mass loss. 
The models have been converted into EEP format using the ISO program \citep{Dotter:2016} to be used as input in METISSE. 
The second set of models were computed using the mass-loss rates from \citet{Belczynski:2010} by modifying the subroutine \verb|run_star_extras| in MESA to match the implementation in METISSE. 
For stars with initial masses of 30\Msun{} and above, the computation of models in the second set also uses MLT++ \citep{Paxton2013} to get around density inversions that hinder the completion of the tracks. Without it, stellar tracks are rendered incomplete due to numerical instabilities during core helium burning \citep[see][for details]{Agrawal:2022b}.

We first compare the evolutionary tracks computed using MESA with and without mass loss, shown in the left panel of Figure~\ref{fig:hrd_metisse_mesa_Zsun}. Differences in the luminosity of the stars are easily visible in the tracks more massive than 10\Msun{}.
The effect of mass loss becomes more pronounced with increasing stellar mass, most notably the change in the position of the main-sequence-hook in stars more massive than 20\Msun{}.
Due to their high luminosity and radius, massive stars can lose up to a few \Msun{} even during the main-sequence evolution. 
The reduced total mass on the main sequence leads to a lower central density and central temperature of the star, which in turn reduces its luminosity and the effective temperature. 
Stars more massive than 30\Msun{} evolve away from the giant branch to higher $T_{\rm eff}$ as mass loss strips off the envelope of the star and exposes the hotter inner layers. 
In the case of a 30\Msun{} model, carbon ignition in the core causes the envelope to expand and cool down, pushing the star back towards the giant phase. 
The 40\Msun{} and 55\Msun{} models lose their envelope completely while undergoing core helium burning and continue their evolution as naked helium stars.

With this understanding of the effect of including mass loss within the MESA models, we next assess how well METISSE can approximate the impact of mass loss on stellar properties. We compare the stellar tracks interpolated by METISSE (labelled as \textit{METISSE-MESA}) using stellar models computed without mass loss by MESA and the mass-loss implementation described in Section~\ref{sec:metisse_mt}, with the stellar models computed with mass loss by MESA (labelled as \textit{MESA-with mass loss}). 
As shown in the right panel of Figure~\ref{fig:hrd_metisse_mesa_Zsun}, the tracks agree well during all phases of evolution for stars with initial masses up to 20\Msun{}. 
For more massive stars, the tracks interpolated by METISSE exhibit a different behaviour compared to the detailed tracks from MESA.
Similar to the detailed MESA tracks, tracks computed with METISSE predict lower luminosities and a change in the position of the MS-hook when mass loss is added in stars above 20\Msun{}. 
However, the degree of change is different and the hook feature does not coincide between the two sets of tracks.  

These differences are further elaborated in Figure~\ref{fig:mass_metisse_mesa}, where we show the evolution of stellar radius, core mass, and total mass of the stars as predicted by the tracks interpolated by METISSE and the detailed MESA models. Similar to Figure~\ref{fig:hrd_metisse_mesa_Zsun}, evolutionary predictions of MESA and METISSE agree very well for stars with initial masses up to 20\Msun{}. 
For more massive stars, METISSE predicts lower stellar radii than MESA towards the end of the main sequence (visible as the small dip in the otherwise gradually increasing radius shown in the top panel of Figure~\ref{fig:mass_metisse_mesa}). 
These differences diminish as the stars expand during the HG and start core helium burning.
In the case of a 30\Msun{} star, MESA predicts a reduction of approximately 1000\Rsun{} in the radius, until carbon ignition in the core causes the stellar envelope to expand to approximately 900\Rsun{}. METISSE predicts a decrease of only about 500\Rsun{} before expanding back to almost the same value as MESA. 

Except for the 55\Msun{} star, MESA and METISSE predict similar values for the maximum radial expansion achieved by the stars during the core helium burning phase. 
Thereafter, MESA predicts a gradual decrease in radius for the 40\Msun{} and 55\Msun{} stars until they lose their hydrogen envelope to become naked helium stars while stars evolved with METISSE maintain the large radii ($\sim$1000\Rsun{}) for tens of thousands of years before losing their envelope and transitioning to SSE formulae for naked helium stars. 
The predictions of stellar radii from the SSE helium star formulae differ significantly from those of MESA and we do not expect agreement in this regime.

The differences in stellar radii during the core helium burning phase and in the naked helium star phase lead to the differences in the evolution of mass between METISSE and MESA (shown in the middle panel of Figure~\ref{fig:mass_metisse_mesa}).
The larger radii predicted by METISSE during core helium burning for the 55\Msun{} star also causes it to lose more mass, resulting in the pre-supernova mass differing by 5\Msun{} relative to the predictions of MESA.  

Discrepancies between the predictions of MESA and METISSE towards the end of the main-sequence phase can also be seen in the predictions of core mass, shown in the bottom panel of Figure~\ref{fig:mass_metisse_mesa}. 
Similar to the predictions of luminosity in Figure~\ref{fig:hrd_metisse_mesa_Zsun} and stellar radius in the top panel of Figure~\ref{fig:mass_metisse_mesa}, METISSE predicts lower helium core masses compared to MESA. 

As shown in the figure, differences in the convective core masses of massive stars only arise at the highest mass-loss rates.
In general, at high mass-loss rates present in massive stars, the removal of the outer layers can destroy its equilibrium.
However, if the mass-loss timescale (time needed to lose the total mass of the star at its current mass-loss rate) exceeds the nuclear timescale (time needed to radiate away star's nuclear energy reservoir at its current luminosity), the star can adjust its structure to the new mass without affecting the core properties \citep{Kippenhahn2012}. 
Towards the end of the main-sequence evolution, the nuclear timescale can be well below the mass loss timescale, and the core evolution can become decoupled from the surface even before the formation of a helium-enriched core.
This is contrary to our SSE-based assumption that the properties of stars during the main sequence are solely determined by the total mass of the star. 
A detailed code like MESA can encapsulate the details of stellar structure, and therefore shows a lower reduction in core properties from the tracks evolved without mass loss compared to METISSE.
In future, the assumptions on how the stellar parameters respond to mass changes in METISSE will be improved to make METISSE mimic detailed evolution as closely as possible.

\begin{figure*}

\begin{tabular}{|c|c|}
        
        \includegraphics[page=1,width=0.33\textwidth]{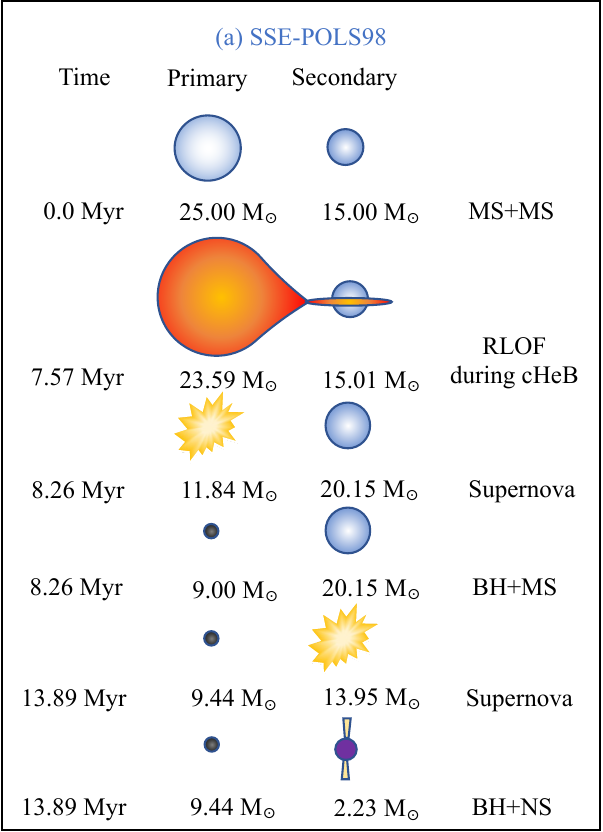}
        \hspace{-0.16in}
&       \includegraphics[page=2,width=0.33\textwidth]{plots/5_plot-crop.pdf}
        \vspace{-0.01in}
\end{tabular}
\begin{tabular}{|l|l|l|}
       
        \hspace{-0.13in}
        \includegraphics[page=3,width=0.33\textwidth]{plots/5_plot-crop.pdf}
        
        \includegraphics[page=4,width=0.33\textwidth]{plots/5_plot-crop.pdf}

        \includegraphics[page=5,width=0.33\textwidth]{plots/5_plot-crop.pdf}
       
\end{tabular}
    
\caption{Evolution summary of a 25\Msun{} and 15\Msun{} binary system of stars with an initial orbital period of 1800\,days and $Z=$ \tento{-2}. Panel (a) represents the evolution of the binary as predicted by BSE-SSE, while other panels (b to e) represent the evolutionary predictions of the same binary with BSE-METISSE using \pols tracks and the three sets of MESA tracks computed with overshooting parameter, $\alpha_{\rm OVS} =0.11, 0.33,0.55$ respectively, as described in Section~\ref{sec:case_study}. 
The asterisk on `RLOF' in panels (d) and (e) denotes that the mass transfer occurred in two or more discontinuous phases before the next point in the summary was reached. }

\label{fig:summary_bse_metisse}

\end{figure*}

\section{Case study: Evolution of a binary system with SSE and METISSE in BSE}
\label{sec:case_study}

Massive stars are commonly found in binaries, and most of them can exchange mass with their companion \citep{Kobulnicky:2007,Sana:2012,Kobulnicky:2014}. 
Uncertainties in their evolution, beyond being important in their own right, present a major hurdle in our efforts to understand the different binary interactions that these stars undergo \citep{DornWallenstein2020, Belczynski:2022, Romagnolo:2022}.

METISSE provides us with the ability to test the impact of these uncertainties in stellar evolution on the lives of massive stars. 
However, we also require a binary evolution mechanism to predict the evolution of binary systems. 
To achieve this, we have integrated METISSE with the binary evolution code BSE.
We provide a brief overview of the physics of mass transfer in binary systems and the treatment of binary evolution in BSE in Section~\ref{sec:mass_transfer_binary} and Section~\ref{sec:bse}, respectively. 

In this section, we use the binary evolution code BSE, first with SSE formulae (BSE-SSE) and then with different sets of stellar tracks interpolated by METISSE (BSE-METISSE) to demonstrate the impact of uncertainties in massive stellar evolution on the formation of gravitational-wave sources.
In particular, we investigate massive, wide binaries similar to the progenitors proposed for double neutron star 
\citep[e.g.,][]{VignaGomez:2018,Chattopadhyay:2019} and neutron star-black hole binaries \citep[e.g.,][]{Chattopadhyay:2020,Broekgaarden:2021}.
As an example, we study the evolution of a massive binary system of stars with initial masses 25\Msun{} and 15\Msun{} at a metallicity $Z=$ \tento{-2} in a circular orbit (eccentricity $e = 0$) with an initial orbital period of 1800\,days.

The binary system is evolved with each of the following methods and models for computing single-star parameters using BSE,

\begin{enumerate}
    \item SSE-POLS98: SSE formulae to the \pols tracks,
    \item METISSE-POLS98: tracks interpolated by METISSE using \pols tracks as input,
    \item METISSE-MESA-SET1: tracks interpolated by METISSE using MESA tracks computed with $ \rm \alpha_{OVS} =0.11$,
    \item METISSE-MESA-SET2: tracks interpolated by METISSE using MESA tracks computed with $ \rm \alpha_{OVS} =0.33$,
    \item METISSE-MESA-SET3: tracks interpolated by METISSE using MESA tracks computed with $ \rm \alpha_{OVS} =0.55$.
\end{enumerate}

Mass loss due to stellar winds is given by \citet{Belczynski:2010} and remnant masses are calculated from \citet{Belczynski:2008} \citep[also `StarTrack prescription' in][]{Fryer:2012}. The maximum neutron star mass is assumed to be 3\Msun{}. 
For a fair comparison between SSE and METISSE, we nullify the kick velocity imparted to remnants during supernova explosions by setting the dispersion factor for the Maxwellian distribution of velocity to zero. 
The full set of input parameters used for both SSE and METISSE are listed in Table\,\ref{tab:bse_input}.

Before proceeding with the evolutionary predictions of the binary system from each set of tracks, we briefly describe the criteria for the stability of mass transfer through Roche-lobe overflow (RLOF) in BSE. 
At any point during the mass transfer through RLOF, if the mass ratio of the donor to the accretor exceeds the critical mass ratio, $q_{\text {crit}}$, then mass transfer is deemed unstable and can result in the formation of common envelope around the binary.  
For donors that are in the giant phase of the evolution (such as the RGB stars or the AGB stars), $q_{\text {crit}}$ is calculated  using \citep{Hjellming:1987,Webbink:1988},

\begin{equation}
q_{\text {crit}}=0.362+\left[3\left(1-M_{\mathrm{c,d}} / M_{\rm d}\right)\right]^{-1},
\label{eq:qcrit}
\end{equation}

where $M_{\rm d}$ and $M_{\rm c,d}$ denote the total mass and core mass of the donor, respectively. For other nuclear-burning phases of the donor, constant values of $q_{\text {crit}}$ from \citet{Webbink:1985} are used.

\begin{table}
\begin{tabular}{p{0.62\columnwidth}p{0.32\columnwidth}}
\cline{1-2}
\textbf{Parameter} & \textbf{Value} \\ \cline{1-2}

Primary Mass (\Msun{}) & 25\\
Secondary Mass (\Msun{}) & 15\\
Metallicity & 0.01 \\
Maximum evolution time (Myr) & 12000 \\
Orbital period (days) & 1800\\
Eccentricity & 0.0\\

Reimers mass loss scaling factor & 0.5\\
Binary wind enhancement & 0.0\\
Helium star mass loss factor & 1.0\\
Common-envelope efficiency parameter & 3.0 \\
Binding energy factor for common envelope  & 0.5 \\

Tides & On \\
White dwarf cooling scheme & \citet{Hurley:2003}\\
Velocity kick at BH formation & Off \\
Remnant mass scheme & \citet{Belczynski:2008}\\
Maximum NS mass (\Msun{})  & 3.0 \\


Dispersion in the Maxwellian for the supernova kick speed (${\rm km\,sec^{-1}}$)& 0.0\\
Wind velocity factor & 0.125\\
Wind accretion efficiency factor & 1.0\\
Bondi-Hoyle wind accretion factor & 1.5\\
Fraction of accreted matter retained in nova eruption & 0.001\\
Eddington limit factor & 10.0\\
Gamma angular momentum loss & -1.0\\

\end{tabular}
\caption{Input parameters for BSE used in testing the implementation of METISSE and comparing the output with SSE. See \citet{Hurley:2002} for an explanation of each input quantity.
}
\label{tab:bse_input}
\end{table}


\begin{figure}
    \includegraphics[width=0.5\textwidth]{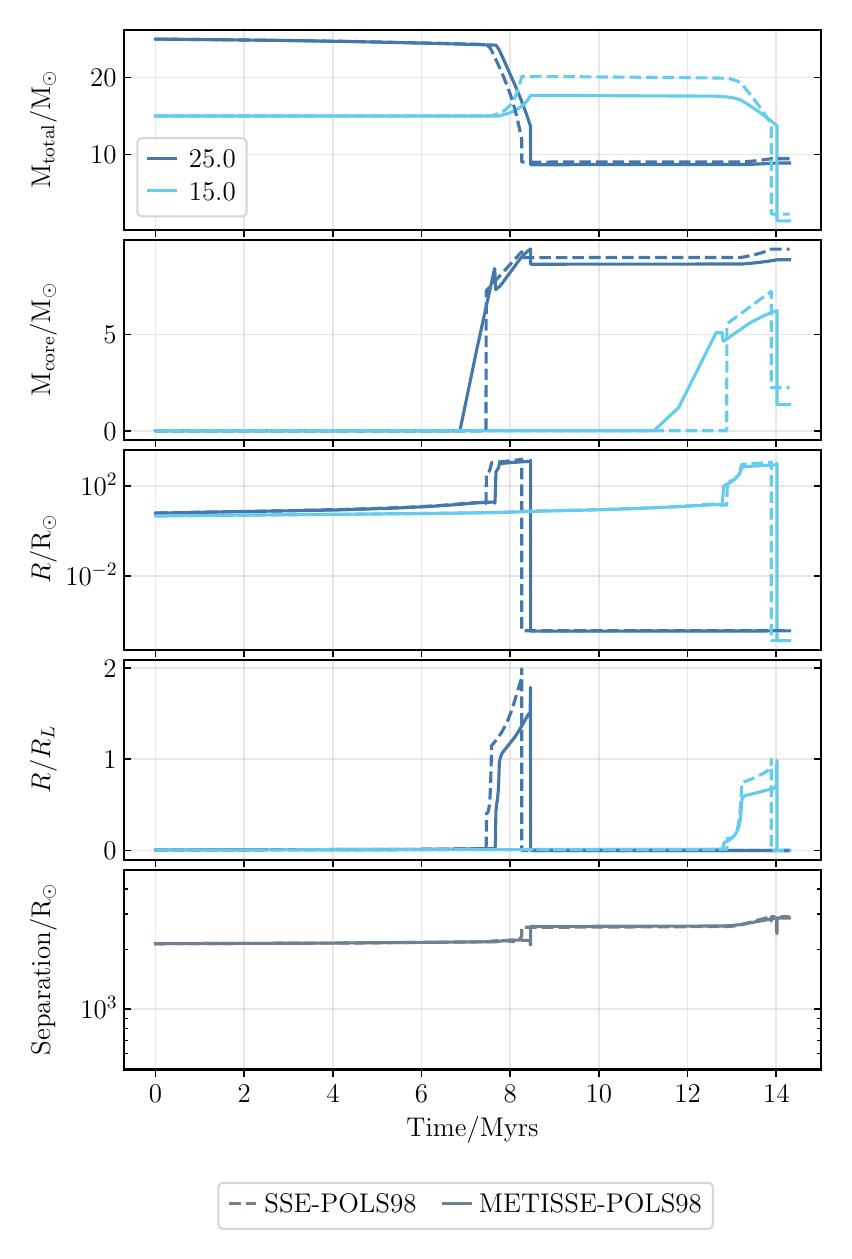}
\caption{Evolution parameters of the binary system illustrated in Fig.~\ref{fig:summary_bse_metisse}, as predicted by BSE-SSE (dashed lines) and BSE-METISSE using \citet{Pols:1998} tracks (solid lines). The dark blue colour represents evolutionary parameters for the 25\Msun{} primary while the light blue colour shows the same for the 15\Msun{} secondary. Due to differences in the radial evolution of the primary between SSE and METISSE, SSE predicts that the primary transfers more mass to the secondary compared to METISSE, resulting in different masses of the remnants.}

\label{fig:comp_bse_metisse}
\end{figure}
\subsection{Binary evolution with SSE and METISSE using Pols98 tracks}
\label{sec:comparing_sse_metisse_with_bse}

In this section, we compute the evolution of the 25\Msun{} and 15\Msun{} binary system described above with BSE-METISSE using \citet{Pols:1998} tracks and compare the results obtained using the traditional method of using SSE formulae in BSE. 
The evolution of the binary system as predicted by both methods is summarized in the top two panels of Figure~\ref{fig:summary_bse_metisse}.  
Panel (a) of Figure~\ref{fig:summary_bse_metisse} depicts the evolution of the binary computed using SSE-POLS98 while panel (b) shows the evolutionary predictions from METISSE-POLS98.

In both cases, the system starts with both stars on the main sequence with an orbital separation of 1800\,days ($\approx$ 2128\Rsun{}). 
The large separation allows the binary to evolve as a detached system, interacting primarily through wind accretion and tides. Some of the mass lost by the 25\Msun{} primary through stellar winds is accreted non-conservatively by the 15\Msun{} secondary. 
As the primary evolves, it expands and eventually fills its Roche lobe during the core helium burning phase, while the secondary is still burning hydrogen on the main sequence.

{RLOF ensues and mass is transferred conservatively from the more massive primary to the less massive secondary.
Mass transfer through RLOF continues on the nuclear timescale as the primary evolves through the giant phases and ends when the primary undergoes a supernova explosion to form a BH.
The secondary, which is now a blue straggler star, continues its evolution while the companion BH grows through wind accretion. The secondary ultimately explodes in a supernova to form a NS. By construction, no natal kick is imparted to the newly formed neutron star, although the large amount of mass lost during the supernova explosion disrupts the binary and the system becomes unbound. }

While the overall evolutionary path for the binary predicted by METISSE agrees well with SSE, there are some discrepancies. 
Both SSE and METISSE predict mass transfer due to RLOF on the nuclear timescale until the primary explodes in a supernova. 
However, the amount of mass transferred from the primary to the secondary with the SSE formulae (5.14\Msun{}) is twice as much compared to METISSE (2.66\Msun{}). 
This is due to the differences in the implementation of mass transfer in both codes during the core helium burning phase of high-mass giant stars (Section~\ref{sec:hrd_metisse_sse}). 
SSE continuously predicts a higher stellar radius compared to METISSE during this phase as the primary loses mass (Figure~\ref{fig:comp_bse_metisse}). 
The ratio of stellar radius to the Roche-lobe radius is similarly higher. 
Since the amount of mass transferred on the nuclear timescale in BSE varies as the cube of the ratio of stellar radius to the Roche-lobe radius \citep[Eqn 58 of][]{Hurley:2002}, more material is transferred from the primary to the secondary in SSE's case as compared to METISSE for a similar amount of time ($\sim$ 0.7\,Myr). 

Another discrepancy can be seen between the METISSE and SSE predictions when each star undergoes a supernova explosion to form a compact remnant (NS or BH). 
SSE, despite predicting a lower pre-supernova mass (11.84\Msun{}) for the primary, predicts the formation of a more massive BH (9\Msun{}) compared to METISSE (8.66\Msun{} BH from a star with pre-supernova mass of 13.6\Msun{}). This discrepancy is a result of the carbon-oxygen core mass predicted by each code: SSE predicts higher a carbon-oxygen core mass of 7.08\Msun{} compared to METISSE (6.68\Msun{}). 
Similarly, despite the similarity between the pre-supernova masses for the secondary ($\sim$ 13.7\Msun{}), SSE predicts the formation of a comparatively more massive NS (2.23\Msun{} compared to NS mass of 1.36\Msun{} by METISSE). 
This is again due to discrepancies in their predictions of carbon-oxygen core mass (5.38\Msun{} compared to 4.35\Msun{} core mass by METISSE).
Since METISSE relies on the underlying input tracks to provide information about the carbon-oxygen core mass, it can better capture the evolution of core. 
SSE, on the other hand, uses a simplified method where the carbon-oxygen core mass grows until a critical value is reached \citep[see Section 5.2 of][for details]{Agrawal:2020}, resulting in a higher carbon-oxygen core mass than METISSE.

\subsection{Binary evolution with METISSE using MESA tracks and different core overshooting}
\label{sec:comparing_binary_evolution_mesa}

Since the publication of the \citet{Pols:1998} stellar models, our understanding of massive stars has expanded considerably. 
In particular, we now have the benefit of several dedicated studies to understand the evolution of massive stars \citep[e.g.,][]{Wade2014MiMes,Szecsi:2016,Renzo:2017,Bjorklund:2021,Grafener:2021}. 
Although many uncertainties remain, the implementation of input physics and parameters used in calculating massive star models have improved significantly over the two decades. For example,
\citet{Pols:1998} used core overshooting calibrated using stars up to 7\Msun{} in binaries and open clusters. 
Since then, \citet{Brott:2011} \citep[also see][]{Castro:2014,Higgins:2019b} have calibrated the overshooting parameter specifically for massive stars using data from the VLT-FLAMES survey \citep{Hunter2008}.

Core overshooting determines the extent to which the convective core of a massive star overshoots in the non-convective outer layers. 
A large value of overshooting implies that more fuel is available to the star during the main sequence as convective mixing brings in unprocessed nuclear fuel. 
Therefore, stars with larger overshooting have a longer main-sequence lifetime as well as a larger helium core mass, radius, and luminosity at the end of it. 
The more massive helium core then undergoes more rapid nuclear burning, resulting in a shorter duration of core helium burning.

In this section, we expand upon our previous analysis of the 25\Msun{} and 15\Msun{} binary system by using MESA models that have been computed using three different core overshooting values as input in METISSE.
These input tracks have been computed using the parameters described in Section~\ref{subsec:metisse_mesa}, using step overshooting and $\alpha_{\rm OVS}= 0.11, 0.33, 0.55$. 
For reference, the value of $\alpha_{\rm OVS}$ used in the tracks from Section~\ref{subsec:metisse_mesa} and by \citet{Brott:2011} is 0.33. \pols used a different method for calculating overshooting and the equivalent value for massive stars is $\approx$ 0.4 in the step overshooting method.

\begin{figure}
    \includegraphics[width=0.5\textwidth]{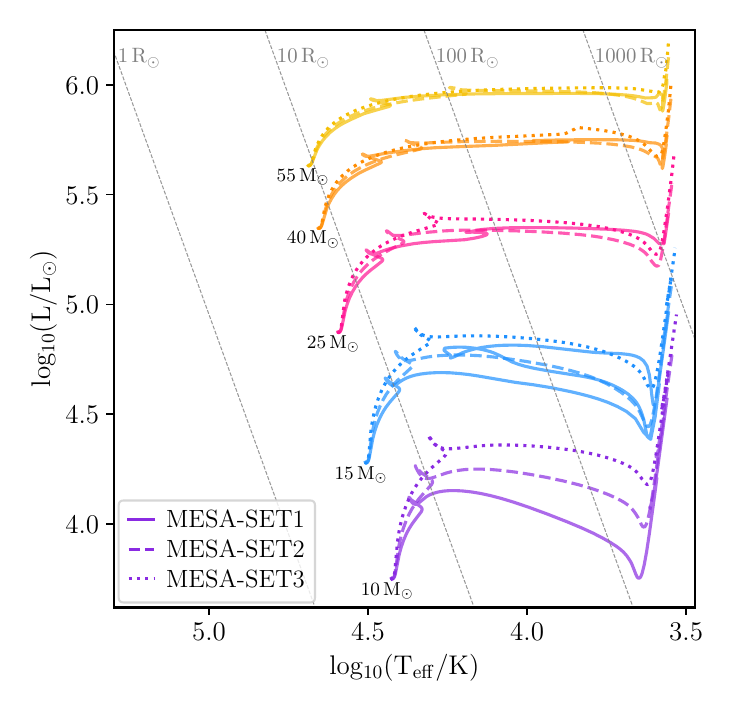}
\caption{HR diagram showing the MESA tracks computed using the same input physics except core overshooting. The three MESA sets have been computed with overshooting parameter, $\alpha_{\rm OVS} =0.11, 0.33,0.55$ and are labelled as MESA-SET1, MESA-SET2, and MESA-SET3 respectively. A larger value of overshooting implies that
more fuel is available to the star during the main sequence, making it larger and brighter.}

\label{fig:HR_MESA_OVS}
\end{figure}

Hereafter we refer to the three sets of MESA tracks as MESA-SET1, MESA-SET2 and MESA-SET3 respectively. 
The evolutionary tracks for a sample of masses from each set are shown in Figure~\ref{fig:HR_MESA_OVS}
For comparison with the \pols models, the input tracks with MESA do not include mass loss and any kind of mass change due to stellar winds or binary interactions has been included using the method described in Section~\ref{sec:metisse_mt}.
The evolution of the 25\Msun{} and 15\Msun{} binary computed using BSE with METISSE and the three sets of MESA tracks as input is depicted in the panels (c), (d) and (e) of Figure~\ref{fig:summary_bse_metisse}.

\begin{figure}
    \includegraphics[width=0.5\textwidth]{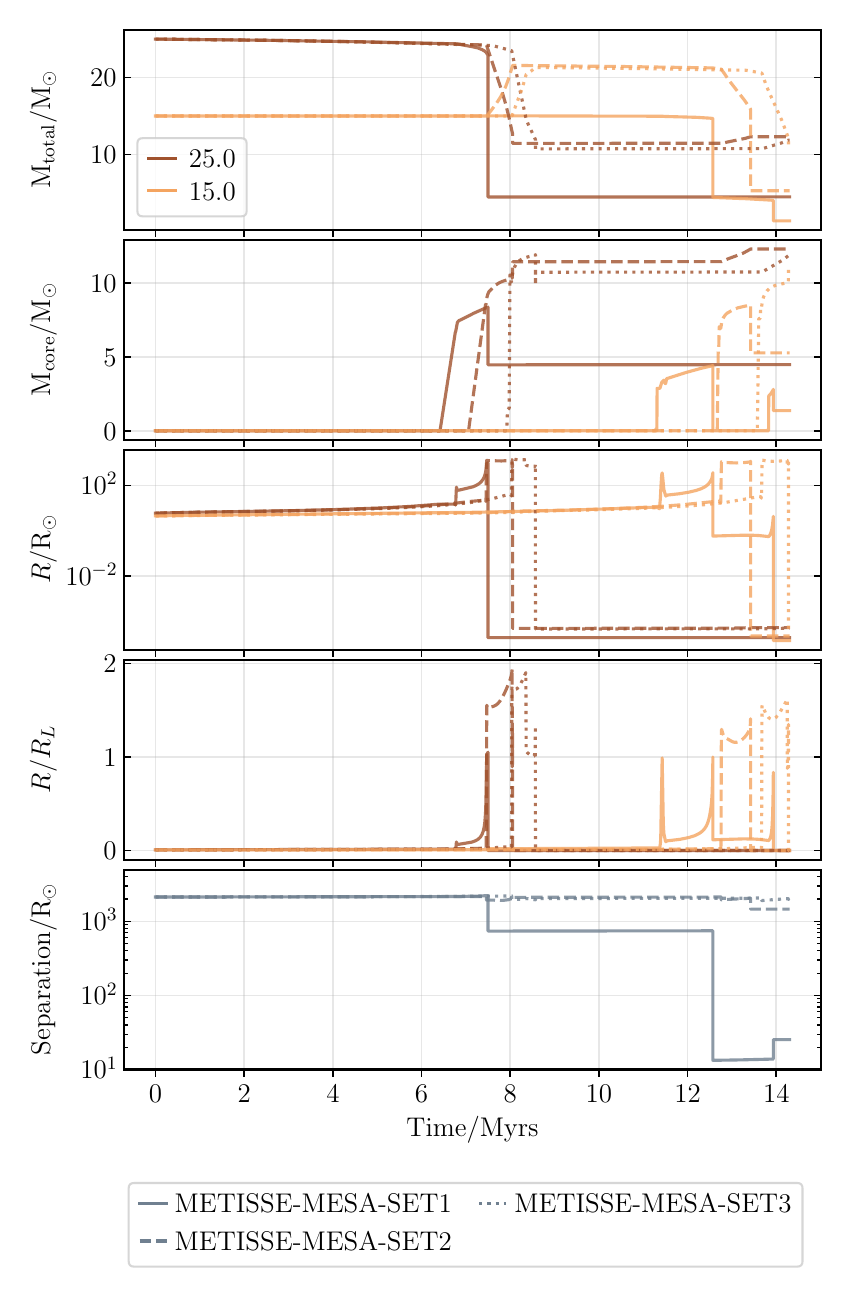}
\caption{Evolution parameters of the binary system shown in Fig.~\ref{fig:summary_bse_metisse}, as predicted by BSE-METISSE with the three sets of MESA tracks computed with $\alpha_{\rm OVS} =0.11, 0.33,0.55$ (labelled as METISSE-MESA-SET1, METISSE-MESA-SET2, and METISSE-MESA-SET3 respectively). Differences in core overshooting can alter the predictions of stellar properties as well as the binary interactions for the same binary system.}

\label{fig:comp_bse_metisse_MESA}
\end{figure}

The evolutionary outcomes in all three cases with MESA tracks differ significantly from one another. 
They are again different from the evolution of the binary computed using the \pols tracks in Section~\ref{sec:comparing_sse_metisse_with_bse}.
In all three cases with MESA tracks, changes in core overshooting changes the stellar radius (Figure~\ref{fig:comp_bse_metisse_MESA}). 
Therefore, RLOF can be initiated at a different time and evolutionary phase in the binary system which, in turn, determines the amount of mass transferred to the secondary. 
For METISSE-MESA-SET1, RLOF starts during core helium burning at a similar primary mass and evolutionary phase as METISSE-POLS98. 
For METISSE-MESA-SET2 and METISSE-MESA-SET3, RLOF begins earlier in the evolution of the binary, when the primary is still on the HG, owing to more overshooting and therefore a larger radius of the primary. 
However, the larger overshooting value also extends the main-sequence life of the star, and RLOF begins at a later time in the case of METISSE-MESA-SET3 compared to METISSE-MESA-SET2.

In the case of METISSE-MESA-SET1, as the primary evolves beyond core helium burning to the AGB phase, the mass transfer becomes dynamically unstable according to Equation~\ref{eq:qcrit}, and a common envelope (CE) forms around the binary. 
Following its CE evolution, the primary loses its envelope to form an 8.37\Msun{} naked helium star with a 6.18\Msun{} carbon-oxygen core that shortly afterwards collapses to form a 4.46\Msun{} BH.
RLOF proceeds differently in case of METISSE-MESA-SET2 where mass transfer remains stable but ends and restarts multiple times due to changes in the stellar radius as the primary evolves beyond the HG to further stages of evolution. 
The primary finally ends its life in a supernova, forming an 11.43\Msun{} BH. 
Owing to a more massive helium core, RLOF in the case of METISSE-MESA-SET3 strips the primary of its hydrogen envelope. 
The primary becomes a naked helium star with a 10.80\Msun{} carbon-oxygen core and eventually ends its life as a 10.72\Msun{} BH.

Similar to the case with \pols tracks (SSE-POLS98 and METISSE-POLS98), all three systems with MESA tracks are left with a BH and MS secondary at the end of primary's life. 
While the mass of secondary remains practically unchanged in the case of METISSE-MESA-SET1, RLOF in the cases of METISSE-MESA-SET2 and METISSE-MESA-SET3 causes the secondary to accrete more than 6 \Msun\ of material and form a blue straggler star. 
In the case of METISSE-MESA-SET1, the common envelope episode reduces the separation between the stars by several hundred \Rsun{}, and mass lost in the supernova explosion of primary hardly increases the orbital separation. 
In the other two METISSE-MESA cases, only about 1\Msun{} is lost in the supernova explosion (compared to more than 3\Msun{} lost in the case with Pols tracks). 
Therefore, unlike the case with the \pols tracks, the orbital separation between the BH and MS secondary remains less than or equal to the original separation of about 2000\Rsun{}. 

This comparatively smaller orbital separation allows the secondary to fill its Roche lobe in each METISSE-MESA case and at a similar evolutionary phase as the primary.  
However for METISSE-MESA-SET1 and METISSE-MESA-SET2, the mass ratio of the binary again renders mass transfer unstable (Equation~\ref{eq:qcrit}) during the core helium burning phase and the system undergoes CE evolution. 
In both cases, the secondary loses its envelope during the CE, forming a naked helium star, but the different core masses cause the secondary to become a 1.35\Msun{} NS in the end in the case of METISSE-MESA-SET1 and a 5.27\Msun{} BH in case of METISSE-MESA-SET2.

A high core mass to total mass ratio keeps mass transfer stable in the case of METISSE-MESA-SET3 (Equation~\ref{eq:qcrit}) and despite reaching a similar mass ratio of primary mass to secondary mass as METISSE-MESA-SET2 during the AGB phase, the binary avoids the CE phase.
The mass transfer remains stable but discontinuous as the star evolves through core helium burning and subsequent phases of nuclear burning.
Finally, the secondary ends its life in a supernova at 14.28\,Myr to form a 10.89\Msun{} BH. 

This example clearly demonstrates how changes to stellar physics, when coupled with complex binary stellar evolution channels, can significantly affect our predictions about gravitational-wave progenitors. The same binary can undergo 0, 1 or 2 CE events depending on the value of overshooting used in the underlying stellar models, remarkably altering the predictions of final separation between the systems.
For METISSE-MESA-SET1, the two CE episodes significantly reduce the separation, leaving a final separation between the two stars of just 25\Rsun{}. 
The end products, a BH and a NS in a tight binary, can eventually merge and can give rise to gravitational-wave emission. 
The single CE phase reduces the orbital separation comparatively little in the case of METISSE-MESA-SET2, ultimately resulting in a wide (1500\Rsun{}) binary black hole system. In the absence of any CE episode in the case of METISSE-MESA-SET3, the orbital separation between the two stars remains quite close to the initial value, 2092\Rsun{}, and the two BHs form a wide binary.

\section{Impact on binary populations}
\label{sec:binary_pop}

\begin{figure*}
    \includegraphics[width=\textwidth]{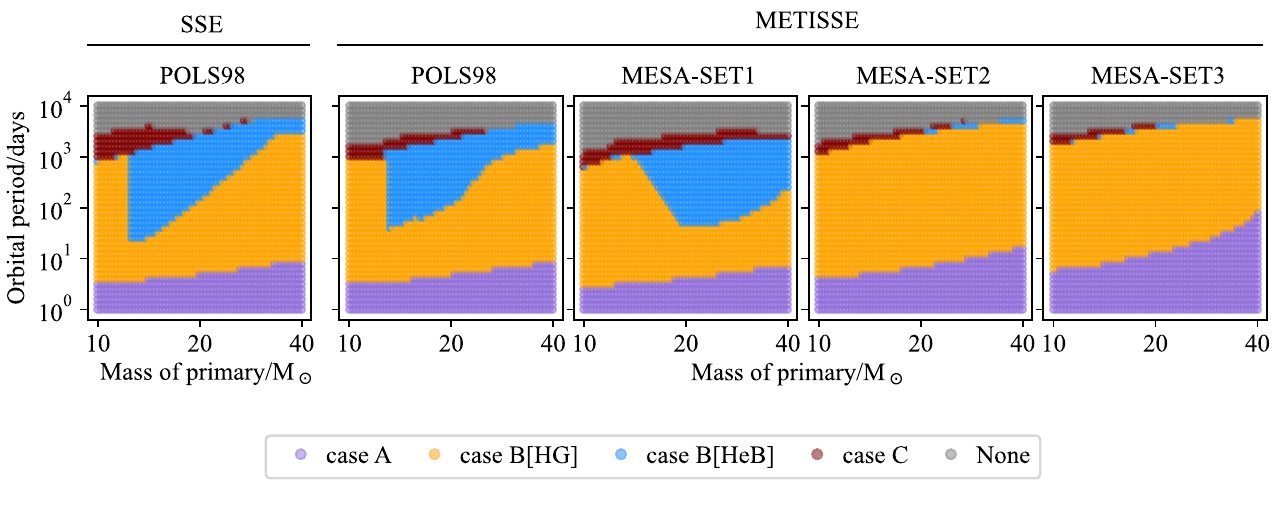}
\caption{Evolutionary state of the donor at the onset of RLOF and the different cases of mass transfer for a population of isolated binaries uniformly distributed in mass and orbital periods as described in Section~\ref{sec:binary_pop}. 
The first panel shows the results when SSE fitting formulae are used for calculating the evolutionary parameters of each star, while the other four panels show the same populations when METISSE is used with \pols{} tracks and the three MESA tracks computed with varying overshooting, for calculating stellar parameters. Similar to Fig.\ref{fig:summary_bse_metisse}, underlying stellar evolution models can play a significant role in determining when RLOF is initiated in the binary, and can potentially affect its final outcome. See Section~\ref{sec:binary_pop} for further details.}

\label{fig:pop_synth}
\end{figure*}

\begin{figure*}
    \includegraphics[width=\textwidth]{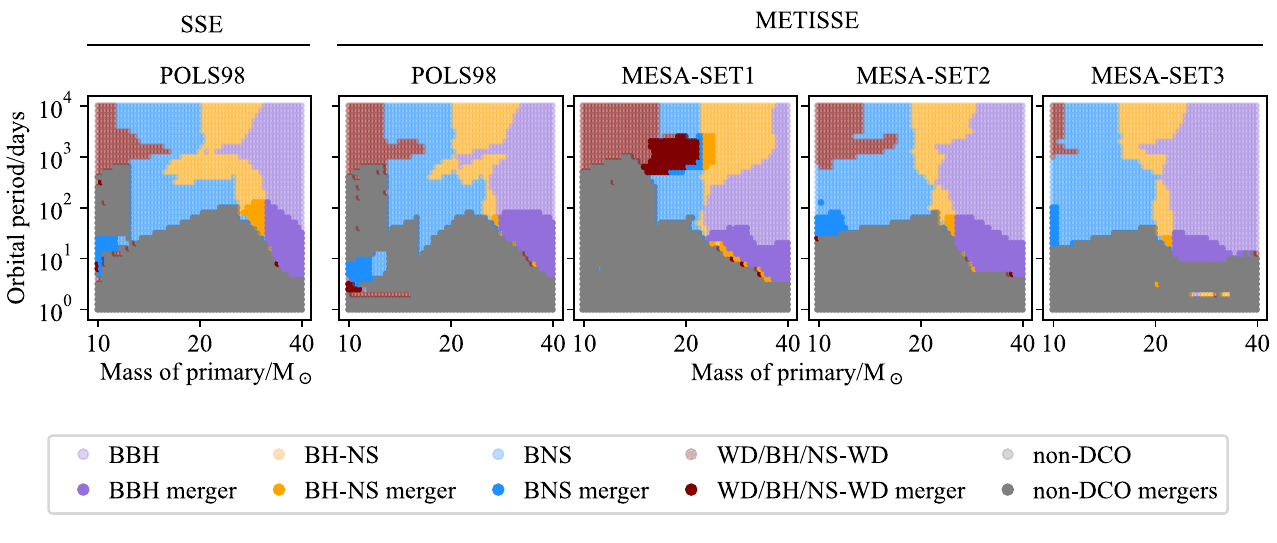}
    \includegraphics[width=\textwidth]{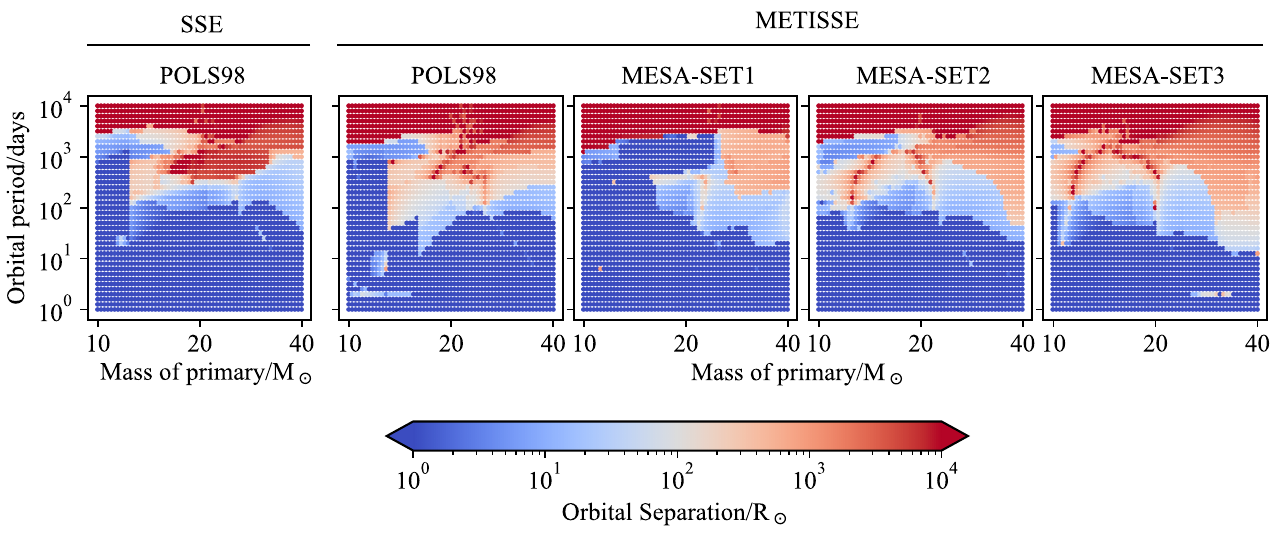}
\caption{Stellar type (top row) and final separation (bottom row) at the end of 12\,Gyr for the different populations of binaries shown in Fig.~\ref{fig:pop_synth}.
A significant fraction of binaries merge before both stars can become a remnant, and are labelled as non-double compact objects mergers (non-DCO mergers; grey circles). Other binaries evolve to form either a binary black hole system (BBH; purple circles), a black hole-neutron star system (BH-NS; yellow circles), a binary neutron star system (BNS; blue circles), or a system containing a compact object with a white dwarf companion (WD/BH/NS-WD; maroon circles)). Darker shades of the same colour represent the systems that merged within 12\,Gyr. 
Based on the underlying stellar models, compact binary mergers, especially BNS mergers, are predicted to occur over a different range of primary masses and orbital periods.
}

\label{fig:pop_synth_rem}
\end{figure*}
Binary stars can interact with each other in a myriad of ways and studying their evolution is an active field of research \citep[e.g.,][]{deMink:2013,Langer:2020,Renzo:2021,VanSon:2022,Temmink:2023}. 
In this section, we extend our analysis of stellar models and their impact on binary evolution to populations of isolated binaries evolved with BSE using the five different configurations for calculating stellar parameters described in Section~\ref{sec:case_study}.
Each population consists of 2501\footnote{The number 2501 is a product of 61 values of initial masses (primary) and 41 values of initial orbital periods.} binaries at metallicity $Z=$ \tento{-2} with primary masses uniformly distributed in log space between 10\Msun{} and 40\Msun{} and spacing of 0.01 dex, and primary to secondary mass ratio, $q = 0.6$. 
All binaries start with a circular orbit (eccentricity, $e = 0$), and orbital periods, $P_{\rm orb}$, ranging from 1 day to \tento{4} days with a logarithmic spacing of 0.1 dex.
The equivalent orbital separation ranges between 10.60\Rsun{} to 4920\Rsun{} for the lowest-mass binary and between 16.80\Rsun{} to 7798\Rsun{} for the most massive binary in the population. 

Mass transfer in a binary occurs when either star fills its Roche lobe and can be classified as one of three cases based on the evolutionary state of the donor star when it first fills its Roche lobe \citep[c.f.,][]{Kippenhahn1967},
\begin{enumerate}
    \item case A: if the donor is undergoing core hydrogen burning on the main sequence,
    \item case B: if the donor has finished core hydrogen burning and is undergoing hydrogen-shell burning (HG, RGB, CHeB),
    \item case C: if the donor has finished core helium burning (AGB).
\end{enumerate}

Depending on the amount and the stability of mass transfer, binary evolution can completely alter the life of the component stars. 
However, until mass transfer through RLOF begins, the evolution of both stars can be approximated by single-star evolution and the uncertainties in the stellar models from which stellar parameters are being calculated are more important than uncertainties in binary evolution. 
The onset of RLOF itself, in particular, depends on the predictions of the radial evolution of the component stars. 
This is evident in Figure~\ref{fig:pop_synth}, where we show the evolutionary state of the donor when the binary experiences mass transfer via RLOF for the first time and the different cases of ensuing mass transfer.

Overall, both SSE-POLS98 and METISSE-POLS98 show very similar morphologies of the different cases of mass transfer except for small variations in case B mass transfer for high-mass stars and case C mass transfer.
The similarities between these two panels can be traced back to their predictions of the radial evolution of the stars (see Section~\ref{sec:hrd_metisse_sse}), where both SSE and METISSE using \pols{} tracks predict a similar radial evolution for stars until the start of core helium burning, and show small variations thereafter. 

In Figure~\ref{fig:pop_synth}, the three METISSE-MESA panels show very different morphologies compared to SSE-POLS98 and METISSE-POLS98, as the different input parameters used in computing the underlying MESA models are quite different. 
With increasing overshooting, predictions of stellar radii also go higher, and RLOF sets in at earlier phases even with larger \porb.
The three sets show variations even between themselves, highlighting the impact of core overshooting, a stellar evolution parameter, in determining the interactions between stars in binaries.
METISSE-MESA-SET1 shows a morphology more closely resembling the METISSE-POLS98 models than the other two MESA sets.
METISSE-MESA-SET2 and METISSE-MESA-SET3, despite appearing similar for much of the parameter space, differ in their predictions of case A mass transfer. 

In all five panels, binaries with \porb{} of up to 3\,days fill their Roche lobe during the main-sequence evolution, leading to case A mass transfer, independent of the mass of the stars or the stellar models in question. 
Since more massive stars have larger main-sequence radii, they undergo case A mass transfer even at larger \porb. 
Similarly, if the stars in the system are too far apart (\porb $>5000$\,days), they never reach radii large enough to fill their Roche lobe and interact only through wind accretion. 

In the case of SSE-POLS98, METISSE-POLS98 and METISSE-MESA-SET1, binaries with primary masses less than approximately 15\Msun  expand to the giant branch before igniting helium in the core, leading to case B mass transfer during the HG. 
A more massive primary ignites helium while on the HG and does not become giant until much later in the core helium burning, leading to case B mass transfer during core helium burning.
Similar to case A mass transfer, the stellar radius on the HG scales with the mass of the primary, and binaries with larger orbital periods can also initiate RLOF while on the HG. 
In the case of METISSE-MESA-SET2 and METISSE-MESA-SET3, larger overshooting causes models to predict larger stellar radii compared to METISSE-MESA-SET1, towards the end of main sequence, and on the HG. Thus, for the same distribution of primary masses and \porb, almost all binaries undergo case B mass transfer on the HG in these two sets.

METISSE-MESA-SET1 further predicts a significant number of binaries with primary masses above 20\Msun{} and \porb{} of about 1000\,days that undergo case C mass transfer. This is again different to the predictions of the other two METISSE-MESA sets (that predict case B mass transfer during the HG), as well as SSE-POLS98 and METISSE-POLS98 (that predict case B mass transfer during core helium burning), for the same distribution of primary masses and \porb.

The timing of the onset of RLOF plays a significant role in determining the evolutionary outcome of the binary. This is clearly visible in Figure~\ref{fig:pop_synth_rem}, where the final properties of the populations are plotted. 
The top row shows the final stellar types for each population whereas the bottom row shows the final orbital separation between the binaries at the end of 12\,Gyr.

The early (case A) onset of RLOF causes a significant transfer of mass and angular momentum between the primary and secondary. 
Binaries with \porb{} of a few days form a contact system and merge before becoming remnants. 
For other binaries with \porb$\lesssim$100\,days, both the case A mass transfer and the case B mass transfer on the HG lead to the reversal of the mass ratio and the binary forms an Algol system, where the secondary is more massive than the primary. 
Further mass transfer then only serves to increase the orbital separation. 
However, the orbital separation remains small enough for the secondary to start RLOF when it evolves off the main sequence. 
By this point, the primary has already become a remnant. 
Due to the extreme mass ratio, mass transfer from the now massive secondary to the remnant is unstable and a CE forms around the binary, drastically reducing the orbital separation.

Binaries with primary masses less than 15\Msun{} and orbital periods between 10\,days and 1000\,days also experience a similar evolutionary path. 
Case B mass transfer on the HG becomes unstable as the primary ascends the red giant branch, leading to CE evolution. 
The CE then reduces the orbital separation such that when the secondary evolves it can initiate the mass transfer through RLOF to the primary that has, by then, become a remnant. 
This mass transfer is also unstable and results in another episode of CE, reducing the orbital separation even further.

In most cases, the successive episodes of CE described in the above cases result in a non-double compact object (non-DCO) merger, i.e., they merge before both stars can complete core nuclear burning to form remnants. 
Due to their shorter nuclear lifetimes, binaries with primary masses more than 25\Msun{}, and orbital periods longer than about 10\,days, avoid merging during the nuclear burning phases and evolve to become binary neutron star (BNS), binary black hole (BBH), or black hole-neutron star (BH-NS) systems. 
In most of these systems, the separation between the remnants is less than 10\Rsun{} and they ultimately inspiral and merge.

In the case of METISSE-MESA-SET2 and METISSE-MESA-SET3, a significant number of binaries with \porb{} longer than 100\,days end up with large final separations ($\gtrsim$ 100\Rsun{}) despite initiating the mass transfer on the HG. 
The larger core masses of these models cause the star to lose its envelope during the stable mass transfer phase before its orbital separation can be reduced significantly. 
Mass loss during the supernova further increases the orbital separation.

The evolutionary outcome of binaries that undergo case B mass transfer during the core helium burning phase depends on whether or not the primary will ascend the giant branch at the end of core helium burning.
Since core helium burning is the second longest phase of a star's nuclear-burning life, the primary can transfer a significant amount of mass to the secondary, forming Algol systems where further accretion widens the binary instead of bringing them closer. 
If the mass transfer continues while the primary evolves to become a giant, it may lead to a CE and reduce the orbital separation.
However, if the mass-loss rate is high, then the primary will not be able to readjust thermally and will move towards lower radii. Thereby ending the mass transfer and leaving behind wide binaries.

At larger \porb{}, unstable case C mass transfer leads to CE events that reduce the orbital separation to a few hundred \Rsun{}. 
Once again, the separation is close enough that when the secondary evolves it undergoes one to two more episodes of CE, further reducing the orbital separation in the range of 10--30\Rsun{}.
Towards the lower mass end, these successive CE episodes also cause the secondary to end up as a WD instead of a NS. 
This manifests as a tail protruding into the BNS region around 15\Msun{} that is visible in all cases.
The impact of these successive episodes of CE is most pronounced in METISSE-MESA-SET1, where a large fraction of binaries that would have ended as BNS systems instead form short-period NS-WD binaries that ultimately merge. Around similar masses and orbital periods, CE evolution also leads to significant numbers of BNS and BH-NS mergers in this set. 
Finally, non-interacting binaries, with \porb{} greater than 5000\,days, eventually get disrupted as mass loss in a supernova explosion during the formation of remnant pushes them further apart.

Each panel also shows significant variations in the region of parameter space (masses and orbital periods) where compact binary mergers are likely to happen. 
At orbital periods less than 100\,days, all three MESA sets predict the formation of a BBH system with primary masses as low as 23\Msun compared to the predictions of SSE-POLS98 and METISSE-POLS98 that require primary masses above 28\Msun{}.
However, METISSE-MESA-SET3, with its larger overshooting and therefore larger core masses, predicts the smallest area in the parameter space where these BBHs can merge. 

All four panels, SSE-POLS98, METISSE-POLS98, METISSE-MESA-SET2 and METISSE-MESA-SET3, predict that binaries with primary masses of 10--12\Msun{} will lead to BNS-mergers, although the orbital periods required for these mergers are different in each case. 
METISSE-MESA-SET1 exclusively predicts the formation of BNS from binaries with primary masses in the range of 15--22\Msun{} and orbital periods varying from 500\,days to 2500\,days.
Although the mergers in this region of parameters space are unique to METISSE-MESA-SET1, they emphasize the role of post-core-helium-burning donors in successful CE ejections, similar to the findings of \citet{Klencki:2021}. 

METISSE-MESA-SET1 predicts BH-NS mergers around similar orbital periods as BNS mergers but at slightly higher primary masses. This is in addition to a small pocket of BH-NS merger around 22\Msun{} and 100\,days that are predicted by all five sets. In this pocket, SSE-POLS98 predicts BH-NS mergers are predicted to occur over a fairly large range of primary masses (26--30\Msun{}) and orbital periods (50--150\,days).
These variations between the five sets further reinforce the importance of stellar parameters in determining the interactions between the binaries, especially compact binary coalescence.

\section{Conclusion and Future Work}
\label{sec:conclusions_chap4}

In this paper, we updated our interpolation-based rapid stellar evolution code METISSE to include the impact of mass changes due to stellar winds and binary interaction.  
We tested the implementation of mass transfer by using the mass-loss rates from \citet{Belczynski:2010} and comparing tracks interpolated with METISSE using stellar models from \citet{Pols:1998} with the tracks computed using the SSE \citep{Hurley:2000} fitting formulae. 
We find that with the current set of assumptions for modelling additional mass loss in METISSE, it can closely reproduce the results from SSE fitting formulae using the same mass-loss rates. 
However, due to the technical details of how mass loss is implemented in both codes, SSE predicts larger radii compared to METISSE towards the end of core helium burning in giant stars more massive than 20\Msun{}.

We further compared METISSE using detailed tracks computed by MESA without any mass loss with the MESA tracks computed with intrinsic mass loss. 
Where the approximation of quasi-static equilibrium holds, METISSE's implementation of mass loss (applied to tracks without mass loss) shows excellent agreement with the tracks from MESA.
Outside this regime, where mass loss is too rapid for the star to maintain equilibrium, METISSE predicts lower luminosity, core masses, and radii compared to MESA. 
The differences are significant only for stars more massive than 30\Msun{}, with the maximum discrepancies being 3\Msun{}, 4\Msun, and 500\Rsun{} in the predictions of pre-supernova core mass, pre-supernova total mass, and maximum stellar radius respectively. 

In future, we will revisit the assumptions underpinning core evolution in the presence of both mass loss and mass gain in METISSE to better mimic the detailed evolution. 
For example, the current implementation of extra mass transfer mechanisms in METISSE also assumes that the structure of the core is unaffected by the mass transfer during the post-main-sequence evolution.
However, recent studies \citep{Renzo:2017, Laplace2021,Schneider:2021} indicate that not only does mass transfer affect the core structure, but the pre-supernova structure of the core is also different for massive single and binary stripped stars. We are also working to include detailed helium star models for stars that lose their hydrogen envelope during nuclear burning (see \citealp{Agrawal:2020} for details).

Overall, there is still a remarkable agreement between the predictions of METISSE and MESA, and the above improvements will only make them better.
While METISSE already had the capability to interpolate between detailed models that included wind mass loss, it now has the capability to add extra mass loss, enabling its use for modelling mass transfer in binary systems.

We utilized this new ability of METISSE to demonstrate the impact of stellar evolution on binary interactions by integrating METISSE with the binary evolution code BSE \citep{Hurley:2002}. 
For an isolated binary system with masses 25\Msun{} and 15\Msun{}, and initial orbital period of 1800\,days, 
BSE predicts similar evolutionary paths for the binary when METISSE is used with \citet{Pols:1998} models for determining single stellar parameters compared to when SSE fitting formulae are used. However, SSE's predictions of larger stellar radii during core helium burning leads to a difference of approximately 2.5\Msun{} in terms of mass transfer from primary to secondary and formation of a 2.23\Msun{} neutron star by the secondary compared to 1.36\Msun{} neutron star predicted by METISSE.

We tested the impact of changing overshooting on the evolution of the above binary by using BSE-METISSE with three sets of MESA models computed with the overshooting parameter, $\alpha_{\rm OVS} =0.11, 0.33,0.55$, and find that differences in stellar properties, especially radii and core masses, can significantly alter the evolution of the binary.
The same binary can undergo different numbers of common envelope episodes depending on the value of overshooting used in the underlying stellar models. The end product can either be a binary black hole system or a black hole-neutron star system with orbital separations ranging between 25\Rsun{} and 2092\Rsun{}. The mass of the remnant formed by the primary ranges between 4.47\Msun{} and 12.30\Msun{}, while the mass of the secondary remnant varies between 1.36\Msun{} and 10.89\Msun{}.

We extended our analysis to populations of isolated binaries uniformly distributed in mass and orbital period. 
The populations have been computed with BSE using SSE fitting formulae, and METISSE's interpolation of \pols{} models as well as the three MESA model sets that have been computed with varying overshooting. 
Similar to the evolution of the isolated binary system, the underlying stellar models play an important role in determining when the mass transfer is initiated and the final evolutionary outcome.
Further, different stellar models suggest different regions of the binary parameter space (in terms of primary mass and orbital period) responsible for compact binary coalescence, especially binary neutron star mergers.
These differences can have important implications for the gravitational-wave merger-rate predictions.

Moreover, the three METISSE-MESA populations highlight the relative impact of varying a single stellar parameter, $\alpha_{\rm OVS}$. 
The difference in binary predictions is much larger between METISSE-MESA-SET1 ($\alpha_{\rm OVS}$ of 0.11) 
and METISSE-MESA-SET2 ($\alpha_{\rm OVS}$ of 0.33) than between METISSE-MESA-SET2 and METISSE-MESA-SET3 ($\alpha_{\rm OVS}$ of 0.55). 
Given the large uncertainties in single-stellar parameters, such comparisons are useful for reducing the size of the parameter space by identifying the regions that have the most influence on binary evolution. 
The analysis can also be repeated with other stellar parameters such as nuclear-reaction parameters or stellar rotation rates.
These comparisons pave the way for codes like POSYDON \citep{Fragos:2022}, which can treat both stellar and binary evolution with more accuracy but require a large set of input models (tens of thousands compared to tens to a few hundred models required by METISSE).

In this work, we limit our analysis to stars at solar-like metallicity, ($Z=$ \tento{-2}). 
Similarly, we have used a constant primary-to-secondary mass ratio, $q = 0.6$, for all binary systems. 
In future, we will be expanding our analysis to lower metallicities and to binaries with different mass ratios.
For comparison with \pols models, we have tested the effect of mass transfer on tracks without any mass loss. 
However, many existing tracks do include wind mass loss and one only needs to account for the effect of binary mass transfer. 
Therefore, we will also be testing the evolution of binaries using models computed with wind mass loss in METISSE in future.

The use of interpolation within METISSE offers great flexibility regarding input stellar models as one can easily use models generated with different stellar evolution codes and different input parameters. 
Similar to other interpolation-based codes such as SEVN and COMBINE, it uses the approach of switching to tracks of different mass in the presence of mass transfer, although the method of finding the new track varies between the codes. 
Moreover, while SSE uses fitting formulae for the stellar parameters that are explicitly required in binary evolution, such as the core radius and envelope binding energy, other interpolation-based codes require them as input data for interpolation. 
METISSE can combine these options freely. Whenever the extra parameters required for binary evolution are present in the input tracks, METISSE interpolates between them. Otherwise, it can switch to using fitting formulae from SSE. 
This approach allows METISSE to be very robust, an important requirement for population synthesis purposes. 

METISSE can be incorporated in binary population synthesis codes such as COSMIC \citep{Breivik2020}, COMPAS \citep{Stevenson:2017,VignaGomez:2018}, and MSE \citep{Hamers:2021} as well as dynamical star cluster modelling codes such as CMC \citep{Rodriguez:2021}, and \textit{N}BODY6 \citep{Aarseth2003} to model diverse astrophysical phenomena in a range of environments, such as the calculations of gravitational-wave merger rates similar to \citet{Broekgaarden:2022}.
Moreover, the modular structure of METISSE and its SSE-like subprogram units (cf. Section~\ref{sec:metisse_as_sse}) for interfacing makes it easier to make modifications in METISSE without affecting the working of the overlying population synthesis codes. This is useful not just for implementing changes related to the current issues with the mass transfer but also for improving METISSE in future. 
Thus, there are a plethora of possibilities with METISSE, and its future looks bright.

\section*{Acknowledgements}

We thank Ross Church, Jan Eldridge, Jakub Klencki, Giuliano Iorio, Michela Mapelli, Katelyn Breivik, for useful comments and discussions.
PA, JH and SS acknowledge support from the Australian Research Council
Centre of Excellence for Gravitational Wave Discovery (Oz-
Grav), through project number CE170100004. This work was supported by NSF Grant AST-2009916 to Carnegie Mellon University and The University of North Carolina at Chapel Hill. CR acknowledges support from a Charles E.~Kaufman Foundation New Investigator Research Grant, an Alfred P.~Sloan Research Fellowship, and a David and Lucile Packard Foundation Fellowship. 
SS is supported by the ARC Discovery Early Career Research Award DE220100241. 
This research was funded in part by the National Science Center (NCN), Poland under grant number OPUS 2021/41/B/ST9/00757. For the purpose of Open Access, the author has applied a CC-BY public copyright
license to any Author Accepted Manuscript (AAM) version arising from this submission

\section*{Data Availability}
The data underlying this article will be shared on reasonable request to the corresponding author.



\bibliographystyle{mnras}
\bibliography{references} 



\appendix

\section{Mass transfer in binary systems}

\label{sec:mass_transfer_binary}

Two stars gravitationally bound to each other and orbiting a common centre of mass are said to form a binary system. 
In the co-rotating reference frame of the binary there exists equipotential surfaces around each star known as `Roche lobes'.
They meet at the inner Lagrangian point ($L_1$), one of five Lagrangian points where the gravitational force exerted by the both stars is balanced by the centrifugal force. The two Roche lobes can be considered to be almost spherical, each with an effective radius known as the `Roche lobe radius'. The Roche lobe radius is defined as the radius of the sphere containing the same volume as the Roche lobe, and is a function of the mass ratio of the stars and the orbital separation between them \citep{Eggleton:1983}.

Most binary systems begin life with each star sufficiently distant that each star's radius is well below its corresponding Roche lobe radius \citep[][]{Abt1983,Duchene2013}.
Even in this detached state, stars can still interact through tides, gravitational radiation, magnetic braking, and through the accretion of stellar winds \citep[][]{Tout:1997,Hurley:2002}.
As the stars expand to become giants and supergiants, their radii can change by several orders of magnitude. 
If one of the stars in the binary system (the donor) expands beyond its Roche lobe radius, its outer layers become dominated by the gravitational influence of the second star (the accretor). 
Material flows through $L_1$, and depending on its energy and momentum, can be accreted by the companion star. 
The binary is now in a semi-detached state and is said to be undergoing mass transfer through RLOF.
A third possibility is that the accretor may fill its own Roche lobe (due to its own stellar evolution or in response to mass transfer on it) and stars form a contact binary system.

Binary interactions can change the mass, angular momentum and the surface composition of the individual stars as well as the properties of the binary system, such as the orbital separation between the stars. 
If the total mass and the total angular momentum of the system is conserved then the mass transfer is described as conservative. In the opposite scenario, both mass and the angular momentum can be lost from the system and the mass transfer is non-conservative. In general, case A and case B mass transfers tend to be conservative while case C mass transfer and wind-accretion are typically non-conservative \citep{Schneider2015}. 

It is also useful to classify mass transfer according to its stability, which depends primarily on the response of the donor stellar radius and Roche lobe radius to mass transfer \citep[see, e.g.,][]{Soberman1997}.
Of critical importance is the response of the donor's envelope to mass loss \citep[][]{Webbink:1985,Ge2015}.
If the donor star has a radiative envelope, it will contract in response to the mass loss, although if the star is on main sequence its nuclear evolution will again lead to an increase in radius. Alternatively, angular momentum losses can lead to a reduction in the Roche lobe radii of the stars. 
In the equilibrium scenario, the donor star stays large enough to just fill its Roche lobe and transfer mass to the companion. The mass transfer is stable and proceeds on the nuclear timescale of the donor star.

For HG stars with radiative envelopes, radial contraction of the donor due to mass transfer can be outpaced by the radial expansion due to stellar evolution.
This leads to increasing mass-transfer rates limited by the thermal time scale of the donor i.e., the time needed to restore the thermal equilibrium of the star.
In this case, mass transfer is unstable and proceeds on the thermal timescale of the donor. 

If the donor star has a deep convective envelope, its radius may increase in response to mass loss, and can quickly lead to a run away situation. The mass-transfer is unstable and can have significant consequences on the evolution of the binary. 
For example, if the expansion of the donor occurs beyond the outer Lagrangian point of the system, or the donor transfers more mass than the accretor can accrete (which is limited by its thermal timescale), then the outer layers of the star form a CE around the binary \citep{Paczynski:1976,Iben1993}. 
As the cores spiral inside the common envelope, their orbital energy is slowly transferred to the envelope, causing subsequent decay of the orbit and may ultimately lead to the ejection of the envelope. If the orbital decay does not lead to stellar merger before the envelope is ejected, it can leave behind two tightly orbiting cores \citep[][]{Ivanova2013}.

Common envelope evolution is a proposed channel for the formation of close binary systems that can lead to phenomena such as Type Ia supernovae, X-ray binaries, and compact binary coalescence. 
In general, most episodes of mass transfer tend to reduce the orbital separation of stellar binaries. However, a binary system can also widen in response to certain mass transfer episodes e.g., when mass transfer occurs conservatively from a lower mass star to a more massive companion. Such a situation can arise, for example, in Algol systems \citep{Plavec1968, Paczynski1971} where an initially more massive donor loses enough mass during its evolution through mass transfer to become less massive than its less evolved companion.

Finally, if a star is massive enough, it will end its life in a supernova, leaving behind a compact object (a NS or a BH) or, in the case of a pair instability supernova, leaving no remnant. Compact objects are thought to receive kicks when they are born \citep{Hansen1997,Hobbs2005}. A combination of mass loss during the supernova explosion and the natal kick received by the newly formed NS or BH may significantly widen or even disrupt the binary \citep[][]{Blaauw1961,Janka1994}.

\section{The binary-star evolution (BSE) code}
\label{sec:bse}

A binary system in BSE is characterized by its metallicity, the mass of the primary (initially more massive) star, the mass of the secondary (initially less massive) star, the orbital period (or orbital separation), and the eccentricity of the orbit. 
The Roche lobe radius for each star is given by the fitting formula from \citet{Eggleton:1983}.
The evolution of the binary is classified into two parts depending on if either of the stars is filling their Roche lobe or not. 
The first part is for detached binary systems where neither of the stars have filled their Roche lobes, and they interact (if at all) solely through wind accretion and tidal interactions. Interaction of detached binaries through tides is treated using parameterized formulae from \citet{Hut1981}, \citet{Zahn1977} and \citet{Campbell1984}, while wind-accretion from a companion star is estimated via the Bondi-Hoyle mechanism \citep{BondiHoyle1944}.

The second part of the evolution is where one or both stars fill their Roche lobes. 
The treatment of RLOF closely follows \citet{Tout:1997}. If the mass transfer becomes unstable according to Equation~\ref{eq:qcrit}, then the binary reaches the state of a CE. The outcome of the CE evolution is determined by comparing the total binding energy of the envelope and the orbital energy of the cores \citep{Paczynski:1976}. The efficiency of energy transfer from the orbit to the common envelope is determined through the parameter $\alpha$ \citep{Webbink:1984} whereas the binding energy of the envelope depends on the structure parameter $\lambda$ \citep{deKool1990}, both of which are treated as free parameters in BSE due to large uncertainties in their values. 

Angular momentum losses through gravitational radiation and magnetic braking are estimated using the weak-field approximation of general relativity \citep{LandauLifshitz1975} and parameterizations from \citet{Rappaport1983} and \citet{Skumanich1972} respectively. If either of the stars undergoes a supernova explosion, a kick velocity is taken randomly from a Maxwellian distribution following \citet{Hansen1997} with dispersion or root-mean-square velocity supplied by the user to calculate the loss in angular momentum from the system.
We refer readers to \citet{Hurley:2002} for further details on binary evolution with BSE. 



\label{lastpage}
\end{document}